\documentclass[aps,reprint,amsmath, amssymb,groupedaddress,floatfix]{revtex4-1}
\usepackage{natbib}
\usepackage{graphicx}
\usepackage{bm}
\usepackage{color,soul}
\usepackage{hyperref}
\usepackage{float}
\usepackage{csquotes}
\usepackage{color,soul}
\hypersetup{colorlinks=true, urlcolor=blue, citecolor=blue}

\begin{document}
\preprint{APS/123-QED}
\title{Second Neighbor Electron Hopping and Pressure Induced Topological Quantum Phase Transition in Insulating Cubic Perovskites}
\author{Ravi Kashikar}
\author{Bramhachari Khamari}
\author{B. R. K. Nanda}
\affiliation{Condensed Matter Theory and Computational Lab, Department of Physics, \\ Indian Institute of Technology Madras, Chennai - 36, India}
\date{\today}
\begin{abstract}

Perovskite structure is one of the five symmetry families suitable for exhibiting topological insulator phase. However, none of the halides and oxides stabilizing in this structure exhibit the same. Through density functional calculations on cubic perovskites (CsSnX$_3$; X = Cl, Br, and I), we predict a band insulator - Dirac semimetal - topological insulator phase transition with uniform compression. With the aid of a Slater-Koster tight binding Hamiltonian, we show that, apart from the valence electron count, the band topology of these perovksites is determined by five parameters involving electron hopping among the Sn-\{s, p\} orbitals. These parameters monotonically increase with pressure to gradually transform the positive band gap to a negative one and thereby enable the quantum phase transition. The universality of the mechanism of phase transition is established by examining the band topology of Bi based oxide perovskites. Dynamical stability of the halides against pressure strengthens the experimental relevance.

\end{abstract}

\maketitle

\section{Introduction}
With the discovery of invariant conducting edge states in insulating CdTe-HgTe-CdTe quantum well\cite{HgTe1,HgTe2}, and conducting surface states in insulating Bi$_{1-x}$Sb$_x$ bulk alloys \cite{Bisb1,BiSb2,BiSb3}, there is a renewed interest in the area of band topology. In this context, using the high throughput means, Yang et al. \cite{nature} have come up with five different symmetry families  that have potential to exhibit topological insulator phase either in their equilibrium structure or under certain mechanical deformations. Four of them are inter-metallic alloys formed by heavier elements (e.g. Bi, Pb, Sb, Te and Se) having stronger spin-orbit coupling (SOC) \cite{bise1,bise2,bise3,bise4} and largely constitute the topological insulator class. The fifth one which is the family of cubic halide perovskites (CsZX$_3$; Z = Sn, Pb, and X = Cl, Br, I) is unique with SOC inactive elements, Cs and X, constitute the major fraction of the chemical composition. All of these compounds are normal insulators (NI) under ambient conditions \cite{ABX1,ABX2}.  
Employing a continuum model, primarily applicable for (Bi, Sb)$_2$(Se, Te)$_3$ family, Freeman et al.\cite{bise1,ABX3} have suggested that by tuning certain interaction parameters, TI phase can be realized in CsPbI$_3$ and CsSnI$_3$\cite{ABX3,ABX4,ABX5}. 

Realization of TI phase in (pseudo)cubic systems, with halogens and oxygen as constituents, will bring several families of compounds, beyond the inter-metallic alloys, into the domain of research on band topology. Furthermore, experimentally it is more feasible to construct heterostructures out of cubic lattice. This will allow to explore rich varieties of emerging phenomena such as Weyl  and  Dirac semi-metal as well as phase transition among them  by synthesizing  TI-TI and TI-Normal insulator interfaces\cite{TI-Interface}. Since, in addition to structural symmetries, the band topology is determined by chemical bonding, bond lengths need to be tuned in order to achieve NI-TI phase transition. Pressure is a natural external stimuli to vary the bond length while maintaining the  cubic symmetry.

In this work, to formulate an universal mechanism that enable the NI-TI phase transition, we have considered the family of halide perovskites CsSnX$_3$ which stabilize 
in the cubic phase at higher temperature (300 - 500 K). The studies are carried out through  DFT calculations, using full potential linearized augmented plane wave method, and parametric diagonlaization of appropriate Slater-Koster tight-binding Hamiltonian. The surface states are estimated using the Wannier formalism. The dynamical stability of these compounds against pressure are verified through phonon studies.  The universal mechanism developed from this study are validated on the family of cubic perovskite oxides ABiO$_3$, where A is a group-I and II element.

\begin{figure}[h]
\centering
\includegraphics[angle=-0.0,origin=c,height=3.5cm,width=8cm]{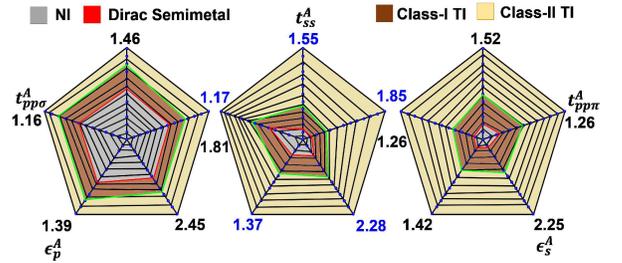}
\caption{
Spider chart illustrating the pressure induced topological phase transition in CsSnX$_3$ in a configure space spanned by five one-electron hopping parameters $t^A$s and $\epsilon^A$s as defined through Eq. 1. The radially increasing equi-pressure contours are shown in black solid lines. The parameters are expressed in terms of their zero-pressure values.
} 
\label{fig1}
\end{figure}

We have identified five interaction parameters, which influence the second neighbor electron hopping among the Sn-\{s, p\} states, account for the band topology of the CsSnX$_3$ family. As sketched in Fig. \ref{fig1}, these parameters increase monotonically with pressure. Beyond a critical pressure, the $s-p$ band inversion occurs at the time reversal invariant momentum (TRIM) to make a transition from NI to TI. At critical pressure the system is an accidental Dirac semimetal\cite{DSM1,DSM2}.

\section{Structural and Computational Details of $CsSnX_3$}

The family of CsSnX$_3$ exhibit different structural
phases at different temperature range. At room temperature, while CsSnBr$_3$ and CsSnCl$_3$ have cubic crystal structures, CsSnI$_3$ has lowered its symmetry and stabilizes in
orthorhombic structure \cite{crystal-cl}. However,  CsSnI$_3$ makes a transition to cubic structure (space group pm-3m) at 425K. \cite{crystal1,crystal2}. In this work, we have examined the  band structure of halide family  in cubic phase, with the lattice parameter listed in Table \ref{T1}.
 To obtain the  band structure, we have carried out DFT+SOC (spin-orbit coupling) calculations through the full-potential linearized augmented plane-wave (FP-LAPW) method \cite{LAPW} as implemented in the WIEN2k package\cite{Blaha}. The generalized gradient approximation (through PBE formalism \cite{GGA}) combined with modified Becke-Johnson (mBJ) exchange potential \cite{mbj-1,mbj-2} is used for the description of exchange-correlation potentials. Augmented plane waves in the interstitial region and localized orbitals inside the muffin-tin sphere are used to construct the basis set. The largest vector in the plane wave expansion is obtained by setting RK$_{max}$ to 7.0. A $12\times12\times12$ k-mesh, yielding 84 irreducible k-points, is used for the Brillouin zone integration.

 Hydrostatic pressure was applied to the experimental cubic lattice to study the change in the band topology. The pressure is calculated using Birch-Murnaghan equation of state\cite{BM1,BM2}.
The bulk moduli as well as the band gaps of the family of CsSnX$_3$ obtained from the DFT+GGA, DFT+GGA+mBJ and GW\cite{ABX2} calculations are listed in Table \ref{T1}. While GW provides a better estimation of band gap, GGA with mBJ correction reproduces the trend and provide reasonable values of E$_g$ to extract the underlying physics of band topology in these materials. 

Ab-initio Molecular Dynamics (AIMD) simulations were performed using VASP simulation package \cite{Kresse,Joubert} to study the temperature dependence of structural stability of the above said compounds. To account for the effect of exchange and correlation, PBE-GGA functional is used\cite{GGA}. A 2$\times$2$\times$2 super cell was considered to create force and displacement data sets. The k-mesh  was set to 2$\times$2$\times$2, with plane wave cutoff energy 500 eV. ALAMODE\cite{Tadano} software is used to get the force constants by fitting force and displacement data.  Phonon band dispersions were obtained by solving the dynamical matrix with given wave vectors.
The phonon band structure of CsSnCl$_3$ with and without compression  are shown in Fig. \ref{fig2}(a, b).  Absence of negative frequencies  suggests the stability of cubic CsSnCl$_3$ under compression. Similar observations are made for CsSnBr$_3$ and CsSnI$_3$. (see Fig. \ref{fig2s} of appendix-B.)

We have calculated the surface states with the use of Wannier formalism. First, the maximally localized Wannier functions (MLWF) as well as the strength of the hopping integral between these functions, are obtained from the bulk DFT calculations using Wannier90\cite{Mostofi}. Taking MLWF as the elements of the basis, a tight-binding model was employed on a slab structure to calculate the surface Green's function through an iterative method as implemented in WannierTools package\cite{arpes}. The imaginary part of the surface Green's function is the local density of state LDOS ($k$, E)\cite{LDOS1, Sancho, Lopez}.

\begin{table}[h]
\centering
\caption{The experimental lattice parameter(a)\cite{ABX1,ABX2}, bulk modulus(B) and its first order pressure derivative (B$^{\prime}$), bulk band gaps (E$_g$) as obtained from DFT+GGA, DFT+GGA+mBJ and GW calculations and experimental studies}. The parameters B and B$^{\prime}$ are instrumental in calculating the pressure as function of volume compression using the Birch-Murnaghan equation of state. 
\begin{tabular}{cccccccc}
\hline
Compound & a(A$^0$) & B (GPa) & B'&  & E$_g$ (eV) &  &  \\
\hline
&  & & & GGA&mBJ&GW&Expt\\
\hline
CsSnCl$_3$ & 5.56 & 25.23 & 2.63 	 & 0.44 & 1.09 &  2.69 & -  \\
CsSnBr$_3$ & 5.80 & 20.83 & 3.09 & 0.04  & 0.41 &  1.38 & 1.75 \cite{crystal2}\\
CsSnI$_3$ & 6.22  & 16.04 & 2.14	 & 0.0  & 0.15 & 1.00 & -  \\
\hline
\hline
\end{tabular}
\label{T1}\\
*The GW values are taken from \cite{ABX2}
\end{table}

\section{Results and Discussion}
\subsection{DFT study}

\begin{figure*}
\centering
\includegraphics[scale = 0.65]{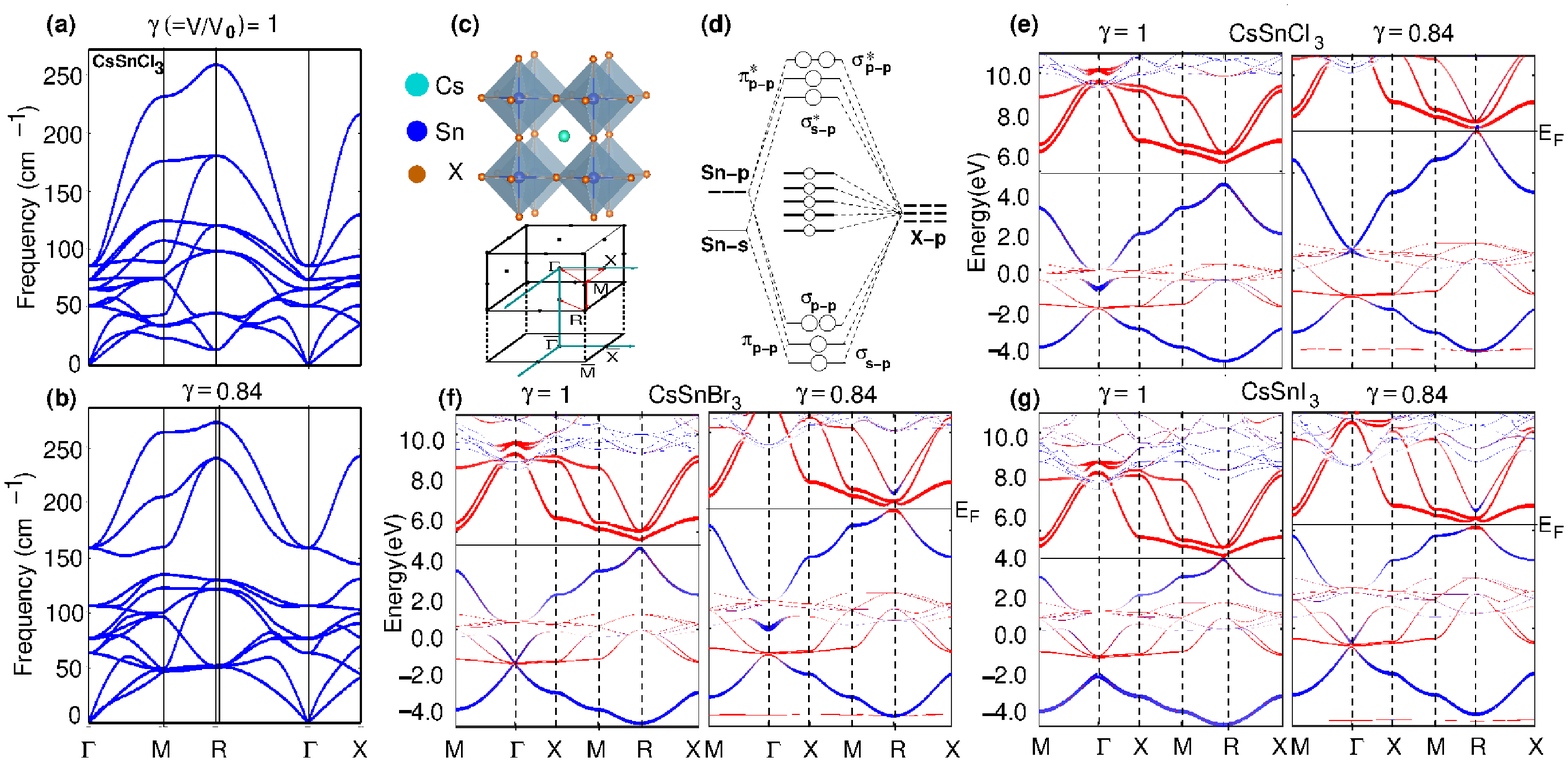}
\caption{Room temperature phonon band structure of CsSnCl$_3$: (a) at zero pressure and (b) at 4.72 GPa ($\frac{V}{V_0} = 0.84$). (c) The cubic lattice with SnO$_6$ octahedra, and the bulk and surface Brillouin zone for CsSnX$_3$. (d) The molecular orbital picture summarizing the effect of nearest neighbor chemical bonding\cite{ABO-1}. (e - g) The band structure of CsSnX$_3$ with and without compression. The compressed structures exhibit s-p band inversion at the TRIM point $R$ ($\pi/a, \pi/a, \pi/a$).
}
\label{fig2}
\end{figure*}

The bulk spin-orbit coupled (SOC) band structure of CsSnX$_3$ at zero-pressure, shown in Fig. \ref{fig2}, provides the following salient features. (i) The Fermi energy (E$_F$) is occupied by four highly dispersive bands. The lower non-degenerate band, henceforth called as singlet, lie below E$_F$ and constitute the valence band maximum (VBM) at $R$. Three other
nearly degenerate bands, henceforth called as triplet, are positioned above E$_F$ and form the conduction band minimum (CBM) at $R$. The gap between VBM and CBM  decreases as we move from X = Cl
(E$_g$ = 1.1 eV) to X = I (E$_g$ = 0.15 eV). 
(ii) A similar set of dispersive bands are seen in the energy window E$_F$ - 8 to E$_F$ - 4 eV creating the band minimum and band maximum at $\Gamma$. They are well separated in the case of I. But in the case of Cl and Br, the band minimum and maximum touch each other.  (iii) A group of five weakly dispersed bands occupy the energy space between these two sets of highly dispersive bands. The band structure of CsSnX$_3$ have a great resemblance with that of the family of perovskite oxides ABiO$_3$\cite{ABO-1, ABO-2, ABO-3}.

The universality in the perovskite band structures can be understood from the molecular orbital picture (MOP) (see Fig. \ref{fig2}(d)) emerging out of stronger nearest-neighbor Sn-\{s, p\} - X-p covalent interactions. As these interactions involve four orbitals of Sn and nine orbitals of three X atoms, they yield a set of four bonding bands, dominated by X-p characters and a set of four antibonding bands, dominated by Sn-\{s, p\} characters. Remaining five bands are nearly flat and are occupied by the X-p electrons. The band gap arises between the singlet and the triplet bands. Depending on the valence electron count (VEC), the gap in the anti-bonding spectrum  either appears at E$_F$ (VEC = 20) or above it. 

In spite of identical MOP of cubic ABiO$_3$ and CsSnX$_3$, there is a significant difference between their band structures. The SOC driven s-p band inversion at $R$, between the lower singlet and upper triplet in the anti-bonding spectrum, is not observed for CsSnX$_3$, while it is present for the family of ABiO$_3$\cite{ABO-1,ABO-2}. Therefore, unlike ABiO$_3$, the family of CsSnX$_3$ do not have topologically protected surface states at ambient conditions. 

Next, we will see if compressing CsSnX$_3$ can introduce band inversion as we may note that the lattice parameter of cubic ABiO$_3$ lies in the range 4.2 to 4.5 \AA \cite{ABX6}, where as for CsSnX$_3$ it is greater than 5.5 \AA. The band structure of CsSnX$_3$ with a compressed volume of 0.84V$_0$, shown in Fig. \ref{fig2}(e - g), suggests the following changes with respect to that of the uncompressed compounds. (i) There is an upward shift in the band energies, as expected, due to a net increase in the repulsive interaction leading to higher eigen energies for one electron states. (ii)The zero-pressure band gap vanishes. (iii) The Sn-s and p dominated antibonding  bands occupying the E$_F$ are more dispersed than the X-p dominated bonding bands, which imply that the second neighbor Sn-\{s, p\} - Sn-\{s, p\} interactions are more sensitive to the pressure.
The VBM and CBM overlapped to create s-p band inversion and the SOC opens a narrow band gap. Since such a gap occurs with overlapping of bands, it is defined as a negative band gap \cite{negative1,negative2, bise1,Bisb1}.

\begin{figure*}
\centering
\includegraphics[scale = 0.5]{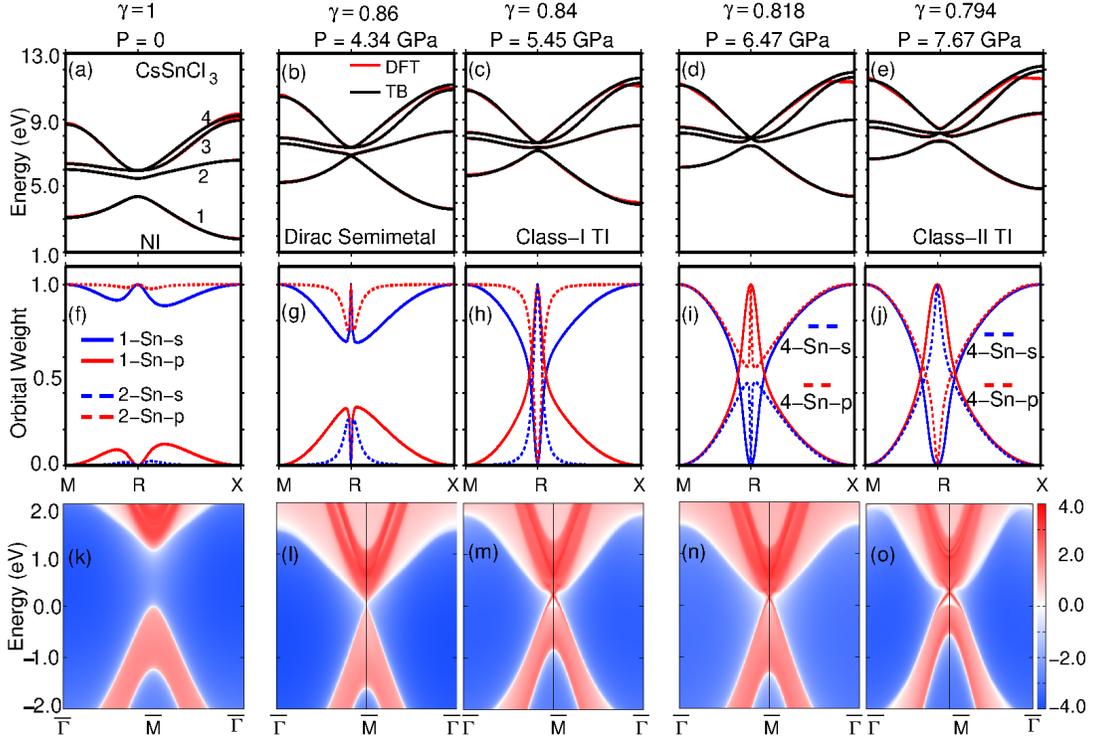}
\caption{(a-e) Bulk SOC band structure of CsSnCl$_3$ in the vicinity of E$_F$ as a function of pressure. (f - j) The Sn-$s$ and $p$ orbital weights of band-1 (solid line) and band-2 or band-4 (dotted line). (k - o) The surface states, calculated using Wannier formalism, are plotted along the path $\Bar{\Gamma}$-$\Bar{M}$-$\Bar{\Gamma}$. Together, the figures show that as we increase the pressure, the bands yield a negative band gap associated with s-p band inversion and thereby create surface Dirac-TI states.}
\label{fig3}
\end{figure*}

\subsection{Tight Binding Model}
 Slater-Koster tight-binding (TB) models are  useful in exploring the underlying quantum mechanical process which occurs due to the structural deformation. While the band structure of CsSnX$_3$ requires a thirteen band TB Hamiltonian involving nine $p$-orbitals of three X atoms, and $s$ and $p$ orbitals of Sn, we have earlier shown that the four antibonding bands in the vicinity of $R$ and E$_F$ are dominated by Sn-\{s, p\} orbitals. Therefore, a four band TB Hamiltonian of Eq. 1, is found to be sufficient to study the band topology of CsSnX$_3$ as a function of pressure. A more technical justification of the appropriateness of this four-band Hamiltonian is presented in the appendix-A. 
\[
H_{TB}
 =  \left( \begin{array}{cc}
    H_{\uparrow\uparrow}& H_{\uparrow\downarrow} \\
    H_{\downarrow\uparrow}^{\dagger}&  H_{\downarrow\downarrow}
\end{array}  \right), 
H_{\uparrow\downarrow}
 =  \left( \begin{array}{cccc}
    0&0&0&0\\
    0&0&0&\lambda\\ 
    0&0&0&-i\lambda\\
    0&\lambda&-i\lambda&0
\end{array}  \right).
\]
\begin{equation}
 H_{\uparrow\uparrow}= (H_{\downarrow\downarrow})^{\dagger}
 =  \left( \begin{array}{cccc}
    \epsilon_s^A+f_0 & 2it_{sp}^AS_x & 2it_{sp}^AS_y & 2it_{sp}^AS_z\\
    -2it_{sp}^AS_x & \epsilon_p^A+f_1 & -i\lambda &0\\ 
    -2it_{sp}^AS_y & i\lambda & \epsilon_p^A+f_2 & 0\\
    -2it_{sp}^AS_z &   0 & 0 & \epsilon_p^A+f_3 
\end{array}  \right)
\end{equation}
\begin{eqnarray}
f_0& = &2t_{ss}^A(cos(k_xa)+cos(k_ya)+cos(k_za)) \nonumber \\
f_1& = &2t_{pp\sigma}^Acos(k_xa)+2t_{pp\pi}^A(cos(k_ya)+cos(k_za)) \nonumber \\
f_2& = &2t_{pp\sigma}^Acos(k_ya)+2t_{pp\pi}^A(cos(k_xa)+cos(k_za))\\
f_3& = &2t_{pp\sigma}^Acos(k_za)+2t_{pp\pi}^A(cos(k_xa)+cos(k_ya))\nonumber
\end{eqnarray}
Here, $\epsilon_s^A$ and $\epsilon_p^A$ are the band center of the singlet and triplet antibonding bands respectively (see Fig. \ref{fig3}(a)), and $\lambda$ is the SOC for p orbitals. While S$_x$ stands for $sin(k_x a)$, the functions $f_0, f_1, f_2$ and $f_3$ are the $k-$ dependent hopping integrals.
The TB bands are fitted with the DFT bands to extract the strength of the hopping parameters and the effective on-site energies. For the zero pressure condition, these values are listed in Table \ref{T2}.  The relative change in these parameters with change in the pressure can be observed from Fig. \ref{fig5}.  From the figure we gather that the effective on-site energy $\epsilon_s$ is very sensitive to the pressure. Concerning the hopping parameters, we find that  compared to t$_{pp\pi}$ and t$_{ss}$ and t$_{pp\sigma}$ vary negligibly with pressure.

\begin{table}
\centering
\caption{ Interaction parameters ($\epsilon^A$s and $t^A$s) and SOC strength $\lambda$ in units of eV at zero pressure. For definitions see Eq. 1 and 3.}
\begin{ruledtabular}
\begin{tabular}{cccccccc}
Compound & $\epsilon_s^A$ & $\epsilon_p^A$ & $t_{ss}^A$ & $t_{sp}^A$& $t_{pp\sigma}^A$ & $t_{pp\pi}^A$ & $\lambda$ \\
 \hline
 CsSnCl$_3$& 2.44 &	7.63 &		-0.32 &	0.47 &	0.73 &	0.1	&0.16\\
CsSnBr$_3$& 2.26 &	6.9 &		-0.31 &	0.49 &	0.8075 &	0.1075 &	0.16\\
CsSnI$_3$ & 2.12 &	6.06 &		-0.23 &	0.45 &	0.8433 &	0.0933 &	0.174
\end{tabular}
\label{T2}
\end{ruledtabular}
\end{table}

The pressure dependent DFT and TB  band structure of CsSnCl$_3$, in the vicinity of E$_F$ and along the path $M - R - X$ are shown in Fig. \ref{fig3} (a - e). The $k-$ dependent orbital contribution for the relevant bands, as obtained from the TB calculations, are plotted in the middle panel. We find that the upward shifting of the singlet band (band-1) due to pressure is large compared to that of the triplet which results in reducing the band gap.  Beyond a critical value of compression, the overlap between band-1 and 2 occurs leading to s-p band inversion and due to the effect of SOC, a new narrow band gap opens up (Fig. \ref{fig3}(c)). Such inversion creates class-I topological insulators as defined for the perovskites\cite{ABO-1}. On further increasing the pressure, nature of band inversion changes and a relatively wider band gap is realized (see Fig. \ref{fig3}(e)). Now the band inversion is between band-1 and 4 which results in formation of class-II topological insulators. We have made similar observations for CsSnBr$_3$ and CsSnI$_3$ and the results are shown in Fig. \ref{fig3s} of appendix-B. 

In Fig. \ref{fig3} (k - o),the local density of state LDOS ($k$, E), which is an imaginary part of the [001] surface Green's function \cite{LDOS1, Sancho, Lopez}, is plotted as a function of pressure for CsSnCl$_3$. 
The colour gradient is a measure of LDOS. Deep red and blue correspond to highest and lowest LDOS. The white color reflects the bulk band structure. The resulted surface states illustrate the phase transition from NI to TI phase with compression.  For the Dirac Semimetal, while the bulk VBM and CBM touches at E$_F$, the surface states create a gap.

\begin{figure}
\centering
\includegraphics[angle=-0.0,origin=c,height=6.5cm,width=7cm]{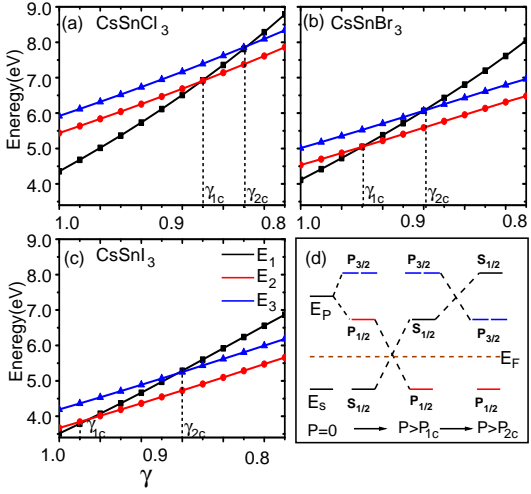}
\caption{
(a) Schematic illustration of orbital manifestation in CsSnX$_3$ under pressure. (b - c) The variation of the energy eigenstates $E_i$(R) of Eq. 3 with pressure for CsSnX$_3$. The crossover between $E_1$ and $E_2$, and between $E_1$ and $E_{3,4}$ are associated with the topological phase transitions. }
\label{fig4}
\end{figure}

The underlying physics of the quantum phase transition from NI to TI can be obtained  from the structure of the Hamiltonian at the TRIM point $R$.  The diagonalization of Eq. 1 at $R$ yields the following eigenvalues.
\begin{eqnarray}
E_1(R) &=& \epsilon_s^A - 6t_{ss}^A \nonumber \\
E_2(R) &=& \epsilon_p^A - 2\lambda - 2t_{pp\sigma}^A - 4t_{pp\pi}^A \nonumber \\
E_3(R) = E_4(R) &=& \epsilon_p^A + \lambda - 2t_{pp\sigma}^A - 4t_{pp\pi}^A 
\end{eqnarray}
The variation of E$_i$(R) with respect to volume compression are shown in Fig. \ref{fig4}. With compression, the bond length decreases and thereby the strength of the interactions increases. 

At zero pressure, $E_2 - E_1$ defines the band gap. With increase in pressure, the gap reduces and at  a critical compression $\gamma_{1c}$, $E_1$ crosses $E_2$ to  form a negative band gap. Therefore, the first band inversion occurs between $s$ orbital dominated $E_1$ and $p$ orbital dominated $E_2$ to form the class-I TI phase. On further compression $E_1$ crosses the double degenerate $E_3$ and $E_4$ at another critical point  $\gamma_{2c}$ to create a new negative band gap to form the class-II TI phase. However, the band inversion is still between $E_1$ and $E_2$ with the doublet $E_3$ and $E_4$ lying in the middle. The compression required to achieve the phase transition is inversely proportional to the zero-pressure band gap. Therefore, while for CsSnI$_3$, the first crossover occurs at $\gamma = 0.974$ for CsSnCl$_3$ it occurs at  $\gamma = 0.866$. As discussed in appendix-D, the E$_g$ is very much dependent on the exchange-correlation functional used in the calculation. However, we find that, the slope of the E$_g$ $\sim$ lattice parameter curve is nearly independent of the exchange correlation functional (see Fig. \ref{fig6s}). Therefore, if the accurate band gap is known for a given lattice parameter, then by using the slope, the band gap for a given pressure,  can be obtained and hence the critical pressures at which the transitions
occur can be estimated with reasonably accuracy.

\begin{figure}[h]
\centering
\includegraphics[angle=-0.0,origin=c,height=7.5cm,width=8cm]{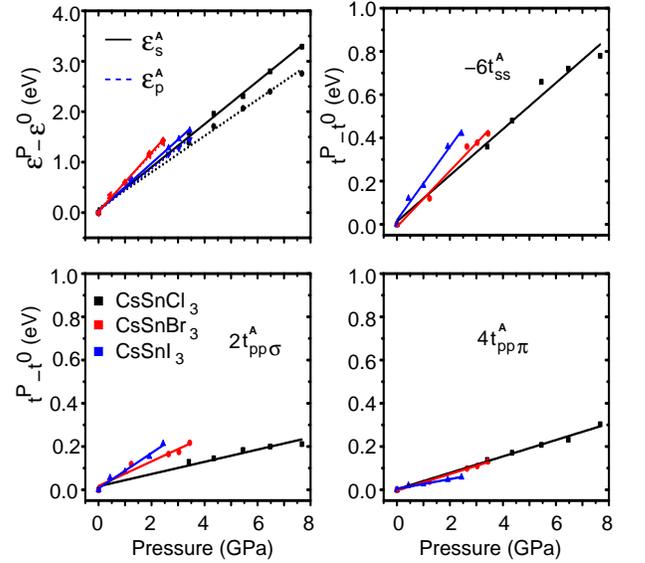}
\caption{Relative increase in effective on-site energies and hopping parameters with increase in hydrostatic pressure. The superscripts \enquote{P} and \enquote{0} represent the values of the parameters at pressure P and at zero pressure (experimental equilibrium structure respectively.)}
\label{fig5}
\end{figure} 

To gain more insight into the physics of topological transition, in Fig. \ref{fig5}, we have plotted each right hand side term of Eq. 3, except $\lambda$,  as a function of pressure. The SOC ($\lambda$) does not depend on pressure.  Through the figure, we reveal that with pressure, the effective on-site parameters, $\epsilon_s$ and $\epsilon_p$, increase nearly equally and hence do not affect the band topology significantly. On the other hand, among the hopping interactions -6t$_{ss}$ increases significantly, compared to the rest, with pressure. Following Eq. 3, as the absolute value of $t_{ss}$ increases, it pushes $E_1$ to cross $E_2$ to induce class-I transition. With further increase in $t_{ss}$, $E_1$ crosses $E_3$ to induce class-II transition.\\

\section{Universality of Phase Transition and Band Structure of $ABiO_3$}
\begin{figure}[h]
\centering
\includegraphics[angle=-0.0,origin=c,height=6cm,width=8cm]{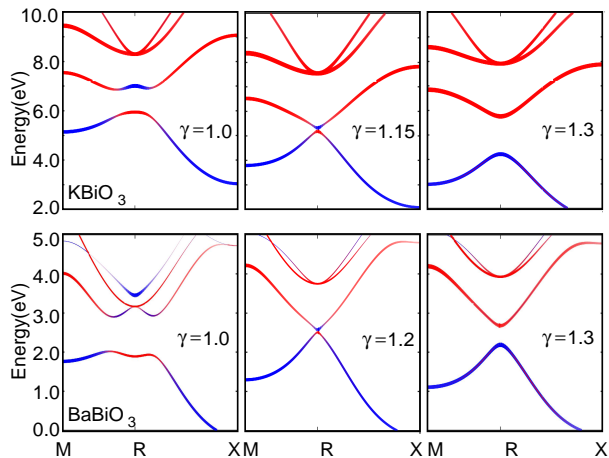}
\caption{Band structure of KBiO$_3$ and BaBiO$_3$ with volume expansion. While the equilibrium structure has s-p band inversion at $R$, with expansion, the inversion vanishes and a normal band gap is realized. Here E$_F$ is set to zero.}
\label{fig6}
\end{figure}
It is desirable to examine the universality of the conditions imposed through Eq. 3 to form pressure induced  symmetry protected TI Dirac states. Since, Eq. 3 is valid only for the cubic perovskite symmetry, we shall examine the family of cubic perovskites ABiO$_3$ to establish the universality. However, unlike CsSnX$_3$,  ABiO$_3$ exhibit the Dirac states in their equilibrium structure\cite{ABO-1}. As mentioned earlier, the equilibrium lattice parameters of the ABiO$_3$ are smaller to that of CsSnX$_3$ approximately by 25\%.   Hence, we shall apply the negative pressure (expansion) and examine the band topology. Fig. \ref{fig6} plots the band structure in the vicinity of $R$ for BaBiO$_3$ and KBiO$_3$ as a function of expansion. With expansion, the non-trivial band gap created through s-p band inversion reduces to form a Dirac semimetal state and on further expansion the trivial band gap appears. The other members of ABiO$_3$ behave identically as can be seen from their pressure-dependent band structure shown in Fig. 12 of the appendix-C.

 To further reconfirm the universality, in Fig. 7, we have plotted the eigenvalues $E_i(R)$, defined in Eq. 3, with respect to relative change in volume $\gamma (= \frac{V}{V_0})$. As in the case of CsSnX$_3$, here also we observe two critical values of $\gamma$ at which the crossover of the eigenvalues occurs. For KBiO$_3$, at the equilibrium structure ($\gamma = 1$), $E_1$ is greater than $E_2$ but lower than $E_3$ to favour the class-I TI phase. Therefore, compression stabilizes the class-II TI phase at the transition point $\gamma_{2C}$. Similarly, with expansion, the NI state stabilizes beyond the transition point $\gamma_{1c}$. However, in the case of BaBiO$_3$, the class-II TI phase ($E_1 > E_3 > E_2$) is observed in the equilibrium structure. Hence, with expansion we observe class-II to class-I transition at $\gamma_{2C}$ and class-I to NI transition at $\gamma_{1C}$. Other members of ABiO$_3$ either follow the trend of KBiO$_3$ or BaBiO$_3$ depending on whether they stabilize in class-I TI phase or class-II TI phase with the equilibrium structure. The results are not shown here to avoid repetition.

\begin{figure}[h]
\centering
\includegraphics[angle=-0.0,origin=c,height=4.55cm,width=8cm]{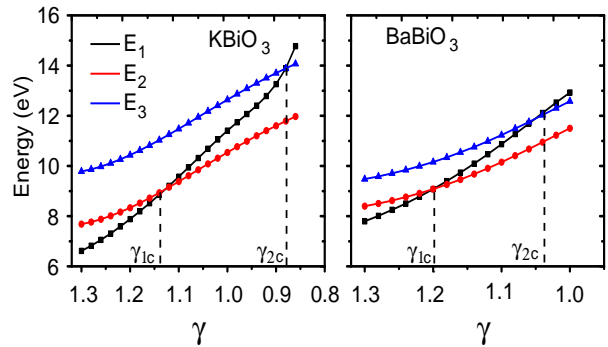}
\caption{Variation of valence band maximum ($E_1$)  and conduction band minimum ($E_2$, $E_3$) at R high symmetry k point of KBiO$_3$ and BaBiO$_3$. At V = V$_0$, KBiO$_3$ shows class-I TI state and BaBiO$_3$ shows class-II TI states. With volume expansion of the unit cell, band inversion disappears making it trivial insulation state.}
\label{fig7}
\end{figure}

\section{Summary and conclusion} To summarize, we carried out electronic structure calculations to  unravel pressure as a controlling parameter to stabilize several electronic phases in cubic perovskite ABX$_3$ stoichiomistry, where A is a group-I and II element, B is a spin-orbit coupled active group-V and VI element and X is either a halogen or oxygen.We show that with pressure, normal insulator to topological insulator quantum phase transition can be achieved and a universal mechanism governs this transition. Furthermore, from the chemcial bonding analysis,  we find that five interaction parameters, namely, $\epsilon_s^A$, $\epsilon_p^A$, $t_{ss}^A$, $t_{pp\sigma}^A$, and $t_{pp\pi}^A$  of Eq. 3, emerging out of  second neighbor electron hopping among the B-\{s, p\} states, determine the band topology in this family. However, the second neighbor B-s - B-s interaction is found to be the driving force for the observed phase transition. The high temperature dynamical stability of these cubic halides against pressure suggests that the phase transition can be realized experimentally.\\

\textbf{ACKNOWLEDGMENT:} The authors acknowledge the computational resources provided by HPCE, IIT Madras. This work is supported by Department of Science and Technology, India through Grant No. EMR/2016/003791.

\appendix

\newpage

\begin{widetext}

\section{Appropriateness of a four band tight binding model}
Beyond the DFT calculations, Slater-Koster tight binding (TB) model \cite{slater} is an useful tool to gain  insight into the electronic structure of crystalline systems. The general TB Hamiltonian  is given by

\begin{equation}
H = \sum_{i,\alpha}\epsilon_{i\alpha}c_{i\alpha}^\dag c_{i\alpha} + \sum_{ij;\alpha,\beta}t_{i\alpha j\beta}(c_{i\alpha}^\dag c_{j\beta} + h.c) + \lambda\textbf{L}\cdot\textbf{S}
\end{equation}
Here, {\it i }({\it j}) and  $\alpha$ ($\beta$ )  are site and the orbitals indices respectively.  The parameters $\epsilon_{i\alpha}$ and $t_{i\alpha j\beta}$ respectively represent the on-site energy and hopping integrals. In the case of CsSnX$_3$, the molecular orbital picture shown in Fig. \ref{fig2}(d) of the main text infers that, a thirteen orbitals basis (one Sn-s, three Sn-p  and nine X-p ) is needed to understand the complete band structure of this family.  

Thus, the spin independent TB Hamiltonian matrix, with the basis set in the order $\{|s^{Sn}\rangle$, $|p^{Sn}_{x}\rangle$, $|p^{Sn}_{y}\rangle$, $|p^{Sn}_{z}\rangle$,  $|p^{X1}_{x}\rangle$, $|p^{X1}_{y}\rangle$, $|p^{X1}_{z}\rangle$, $|p^{X2}_{x}\rangle$, $|p^{X2}_{y}\rangle$, $|p^{X2}_{z}\rangle$, $|p^{X3}_{x}\rangle$, $|p^{X3}_{y}\rangle$, $|p^{X3}_{z}\rangle$$\}$, 
can be written as

\begin{equation}
H = \left( \begin{array}{cc}
M_{4\times4}^{Sn-Sn}&M_{4\times9}^{Sn-X}\\\\
(M_{4\times9}^{Sn-X} )^{\dag}&M_{9\times9}^{X-X}
\end{array}  \right)
\end{equation}.

The individual blocks of this matrix are as follows,

\begin{equation}
M_{4\times4}^{Sn-Sn} = \left( \begin{array}{cccc}
\epsilon_s+g_{1}	& 2it_{sp\sigma}^{Sn-Sn}\sin(k_xa)& 2it_{sp\sigma}^{Sn-Sn}\sin(k_ya) &2it_{sp\sigma}^{Sn-Sn}\sin(k_za)  \\
-2it_{sp\sigma}^{Sn-Sn}\sin(k_xa)	&\epsilon_{p1}+g_2 &	0	&0	\\
-2it_{sp\sigma}^{Sn-Sn}\sin(k_ya) &0&\epsilon_{p1}+g_3&0\\
-2it_{sp\sigma}^{Sn-Sn}\sin(k_za)	&0	&0	&\epsilon_{p1}+g_4 \\
\end{array}  \right)
\end{equation}

\begin{equation}
M_{4\times9}^{Sn-X} = \left( \begin{array}{ccccccccc}
t_{sp\sigma}^{Sn-X}S_x 	&0	&0	&0	& t_{sp\sigma}^{Sn-X}S_y	&0	&0	&0	& t_{sp\sigma}^{Sn-X}S_z \\
t_{pp\sigma}^{Sn-X}C_x	&0	&0	&t_{pp\pi}^{Sn-X}C_y   &	0	&0	&t_{pp\pi}^{Sn-X}C_z&	0	&0\\
0&t_{pp\pi}^{Sn-X}C_x&0	&0&t_{pp\sigma}^{Sn-X}C_y&0 &0&t_{pp\pi}^{Sn-X}C_z &0\\
0	&0	&t_{pp\pi}^{Sn-X}C_x	&	0&0&t_{pp\pi}^{Sn-X}C_y 	&0	&0	&t_{pp\sigma}^{Sn-X}C_z\\
\end{array}  \right)
\end{equation}
Analyzing the partial density of states obtained from the density functional calculations, we have found that, X-p dominated bands are very narrow $(< 1.0$ eV), suggesting negligible X-\{p\}-X-\{p\} second interactions. Hence the block $M_{9\times9}^{X-X}$ can be approximated as

\begin{equation}
M_{9\times9}^{X-X} = \left( \begin{array}{ccccccccc}

\epsilon_{p2}&0&0&0&0&0&0&0&0\\
0&\epsilon_{p2}&0&0&0&0&0&0&0\\
0&0&\epsilon_{p2}&0&0&0&0&0&0\\
0&0&0&\epsilon_{p2}&0&0&0&0&0\\
0&0&0&0&\epsilon_{p2}&0&0&0&0\\
0&0&0&0&0&\epsilon_{p2}&0&0&0\\
0&0&0&0&0&0&\epsilon_{p2}&0&0\\
0&0&0&0&0&0&0&\epsilon_{p2}&0\\
0&0&0&0&0&0&0&0&\epsilon_{p2}

\end{array}  \right)
\end{equation}

Here, $\epsilon_s$, $\epsilon_{p1}$ and $\epsilon_{p2}$ are on-site energies of Sn-s, Sn-p and X-p orbitals respectively. The terms $C_x$ and $S_x$ are short notations for $2\cos(k_xa/2)$ and $2i\sin(k_xa/2)$ respectively. The dispersion term  $g_i$ $(i=1,2,3,4)$, arising from Sn-Sn second neighbour interactions,  are given by
\begin{eqnarray}
g_1 &=& 2t_{ss}^{Sn-Sn}(cos(k_xa)+cos(k_ya)+cos(k_za)) \nonumber \\
g_2 &=& 2t_{pp\sigma}^{Sn-Sn}cos(k_xa)+2t_{pp\pi}^{Sn-Sn}[cos(k_ya)+cos(k_za)] \nonumber \\
g_3 &=& 2t_{pp\sigma}^{Sn-Sn}cos(k_ya)+2t_{pp\pi}^{Sn-Sn}[cos(k_xa)+cos(k_za)]\\
g_4 &=& 2t_{pp\sigma}^{Sn-Sn}cos(k_za)+2t_{pp\pi}^{Sn-Sn}[cos(k_xa)+cos(k_ya)] \nonumber
\end{eqnarray}

The analytic expression for eigen values  at time reversal invariant momentum (TRIM) R $(\frac{\pi}{a}, \frac{\pi}{a}, \frac{\pi}{a})$ are obtained as

\begin{align}
E_1[1]& = \frac{(E_p^X + E_s^{Sn})}{2} -3t_{ss}^{Sn-Sn} - \frac{\sqrt{(E_p^{X} - E_s^{Sn} + 6t_{ss}^{Sn-Sn})^2 + 48(t_{sp}^{Sn-X})^2}}{2} \\
E_2[8]& = E_p^{X}\\
E_3[3]& = E_{p}^{Sn} - 2t_{pp\sigma}^{Sn-Sn} - 4t_{pp\pi}^{Sn-Sn}\\
E_4[1]& = \frac{(E_p^X + E_s^{Sn})}{2} -3t_{ss}^{Sn-Sn} + \frac{\sqrt{(E_p^{X} - E_s^{Sn} + 6t_{ss}^{Sn-Sn})^2 + 48(t_{sp}^{Sn-X})^2}}{2}    
\end{align}

The number in square bracket indicates the degeneracy of the eigenvalues. As a case study, the thirteen band model is applied to CsSnI$_3$ and corresponding bands are fitted with DFT bands to obtain on-site and hopping interactions which are  listed in Table \ref{T3}. The DFT and TB bands are shown in Fig. \ref{fig1s}, suggesting the excellent agreement. 

\begin{figure}[h]
\centering
\includegraphics[angle=0.0,origin=c,height=7cm,width=6cm]{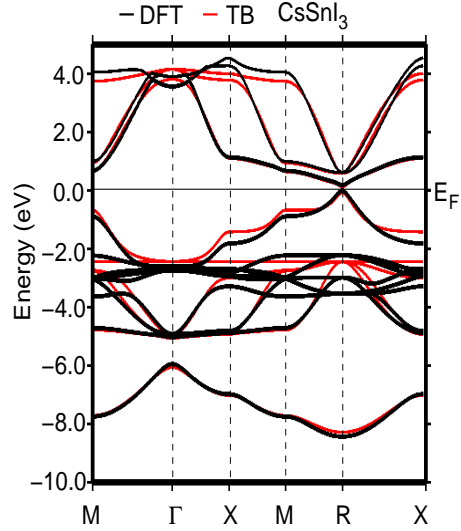}
\caption{ Full basis TB band structure of CsSnI$_3$ and DFT band structure.  }
\label{fig1s}
\end{figure}

\begin{table}[h]
\centering
\caption{ On-site and hopping parameters of CsSnI$_3$ in units of eV}
\begin{tabular}{ccccccccccc}
\hline
$E_{Sn-s}$ &$E_{Sn-p}$ &$E_{X-p}$&$t_{sp}^{Sn-X}$ &$t_{pp\sigma}^{Sn-X}$ & $t_{pp\pi}^{Sn-X}$& $t_{ss}^{Sn-Sn}$ & $t_{sp\sigma}^{Sn-Sn}$&$t_{pp\sigma}^{Sn-Sn}$&$t_{pp\pi}^{Sn-Sn}$ & $\lambda$\\
 \hline
 -5.99	&0.95&	-2.45&	-1.08&	1.9&	-0.5	&-0.014& -0.2&0.239&0.014&0.16 \\
 \hline
\end{tabular}
\label{T3}
\end{table}

The bands which are forming the VBM and CBM at TRIM point are primarily formed by the four Sn-\{s, p\} orbitals. From our calculations we have found that, the contribution to each of these bands by the Sn-\{s, p\} orbitals is more than 70\%. This is true for CsSnCl$_3$ and CsSnBr$_3$ as well. Since, the band topology of CsSnX$_3$ are determined at R, it is prudent to construct a minimal basis set TB Hamiltonian involving these four anti-bonding bands. The SOC incorporated four band TB Hamiltonian in the matrix form can be written as    

\begin{equation}
H= \left( \begin{array}{cccccccc}

    \epsilon^A_s+f_0 & S_x & S_y & S_z & 0 &  0 & 0 & 0\\
    S_x & \epsilon^A_p+f_1 & -i\lambda & 0 &  0 &  0 & 0 & \lambda \\
    S_y & i\lambda & \epsilon^A_p+f_2 & 0 &   0 &  0 & 0 & -i\lambda\\
    S_z &   0 & 0 & \epsilon^A_p+f_3 &    0 &  \lambda & -i\lambda & 0\\
    0 & 0 & 0 & 0 &          \epsilon^A_s+f_0 &  S_x & S_y & S_z\\
    0 & 0 & 0 & \lambda &    S_x &    \epsilon^A_p+f_1 & i\lambda & 0\\
    0 & 0 & 0 & i\lambda &   S_y &    -i\lambda & \epsilon^A_p+f_2 & 0\\
    0 & \lambda & i\lambda & 0 & S_z &     0 & 0 & \epsilon^A_p+f_3

\end{array}  \right)
\end{equation}
Here, $\epsilon^A$s are band centers of the anti-bonding bands and $t^A$s are the second neighbour hopping integrals. The dispersion functions $f_i$ ($i = 1, 2, 3, 4)$ are expressed in Eq. 2. The TB bands obtained from this four band model are shown in the Fig. \ref{fig3} of the main text, and they completely match with DFT bands in the vicinity of the TRIM point R.

\section{Effect of Hydrostatic Pressure on CsSnBr$_3$ and CsSnI$_3$ Band Structure}
As discussed in the main text, the normal insulator (NI) - topological insulator (TI) phase transition can achieved by applying hydrostatic pressure. Fig. \ref{fig2s} shows the phonon band structure of cubic CsSnBr$_3$ (at T = 300K) and CsSnI$_3$ (at T = 425K)  with and without compression. The absence of  negative frequencies proves the dynamical stability of these compounds in cubic phase.

\begin{figure}[h]
\centering
\includegraphics[angle=0.0,origin=c,height=8cm,width=10cm]{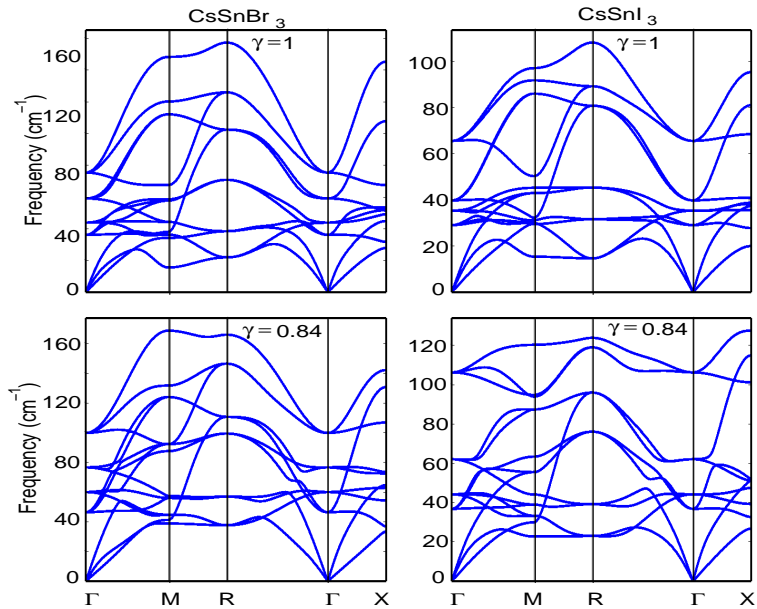}
\caption{ Phonon band dispersions of CsSnX$_3$ (X=Br, I)}.  Here, $\gamma = \frac{V}{V_0}$ 
\label{fig2s}
\end{figure}

Like CsSnCl$_3$ (see Fig. \ref{fig3} of the main text), the CsSnBr$_3$ and CsSnI$_3$ also undergo NI-TI phase transition. The Fig. \ref{fig3s}, where we have plotted bulk and surface band structure as a function of pressure, establishes this phase transition.  

\begin{figure}[H]
\centering
\includegraphics[angle=0.0,origin=c,height=13cm,width=13cm]{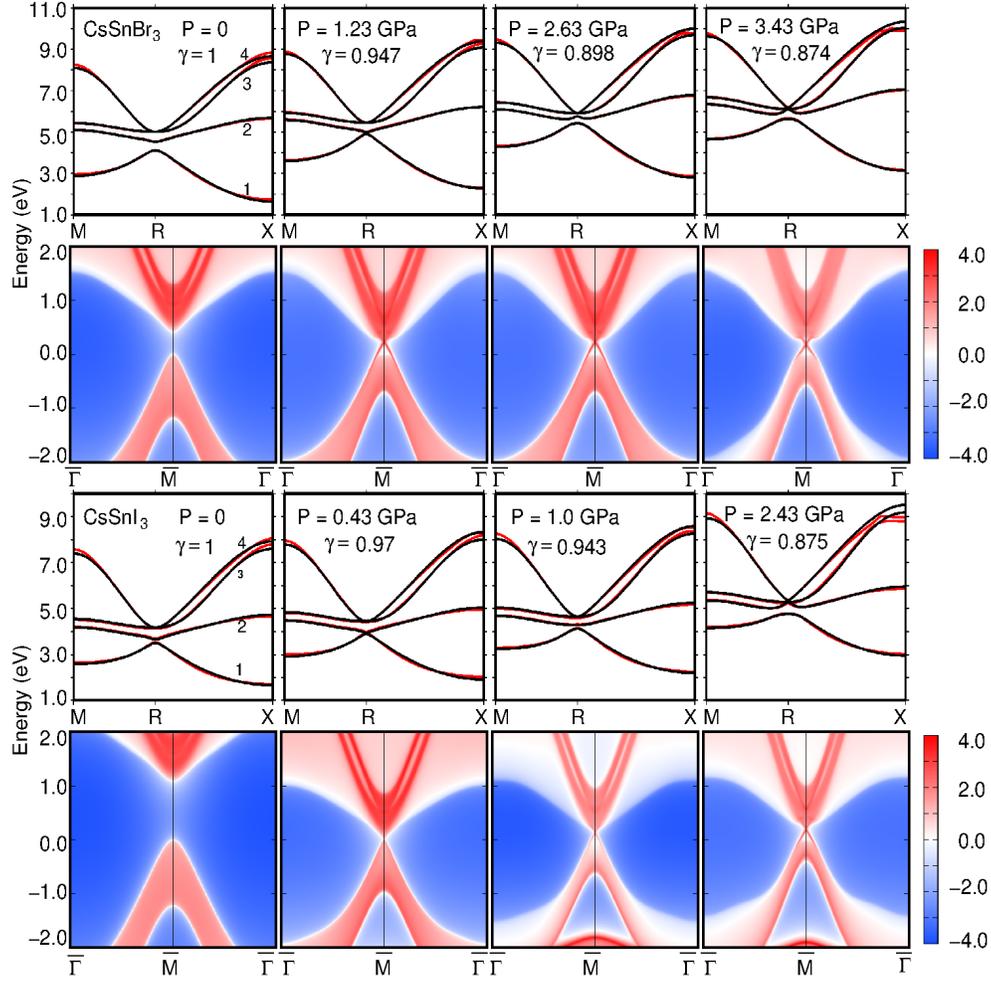}
\caption{ First and second row show the bulk and surface band structure of CsSnBr$_3$ respectively. Third and fourth row shows the same for CsSnI$_3$. With pressure the positive band gap transforms to a negative band gap, which induces s-p band inversion. As a consequence topologically protected surface states are formed at  point $\Bar{M}$ (whose equivalent point is R in bulk Brillouin zone). For details see the main text.} 
\label{fig3s}
\end{figure}

Fig. \ref{fig4s} shows the band gap E$_g$ and applied pressure P as a function of volume compression $\gamma$. The value of  pressure is obtained using Birch-Murnaghan equation\cite{BM1,BM2}. The yellow and white regions identify the NI and TI phase respectively. The border between these two phases stabilizes the  Dirac semimetal state.  Owing to the large bulk band gap CsSnCl$_3$ requires more pressure to induce the NI-TI phase transition. Similarly due to narrow band gap CsSnI$_3$ requires less pressure for the same.   

\begin{figure}[h]
\centering
\includegraphics[angle=0.0,origin=c,height=4.5cm,width=15cm]{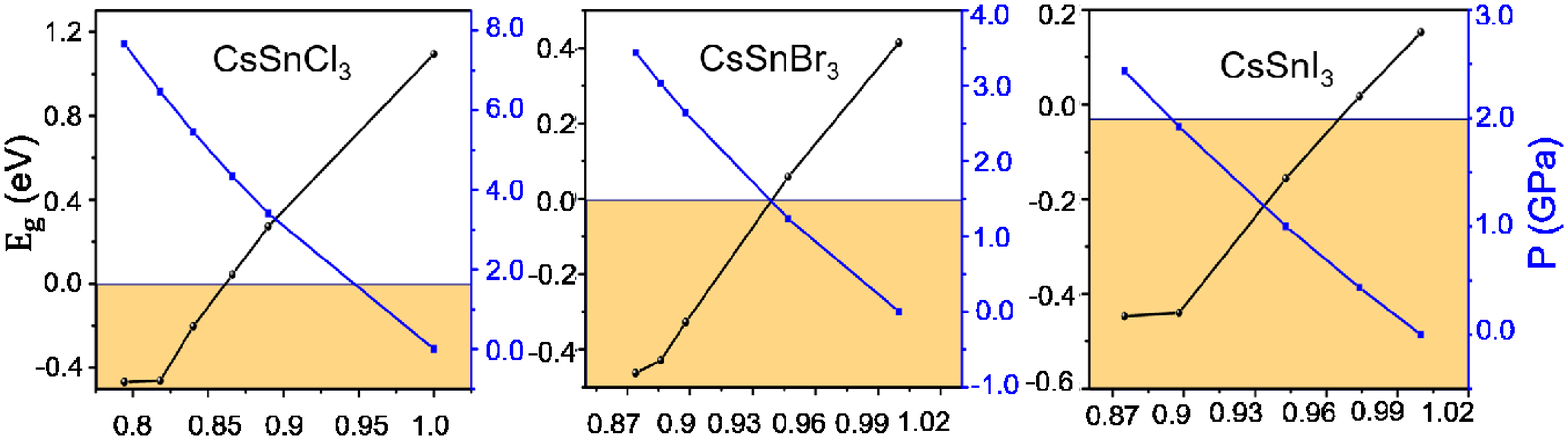}
\caption{Variation of energy gap and pressure with volume compression of different halide perovskites.}
\label{fig4s}
\end{figure}
\section{TI-NI phase transition in $ABiO_3$ }
In section-IV, we have made a discussion about how there is a universal mechanism that governs pressure induced NI-TI phase transition. To further reconfirm the universality of the transition phenomena, in Fig. \ref{fig5s}, we shown pressure dependent band structure of four more perovskites, namely, CsBiO$_3$, NaBiO$_3$, RbBiO$_3$ and SrBiO$_3$. The figure shows that, with expansive pressure, the non-trivial band gap mediated by s-p band inversion vanishes and a normal band gap appears.   

\begin{figure}[h]
\centering
\includegraphics[angle=0.0,origin=c,height=16cm,width=13cm]{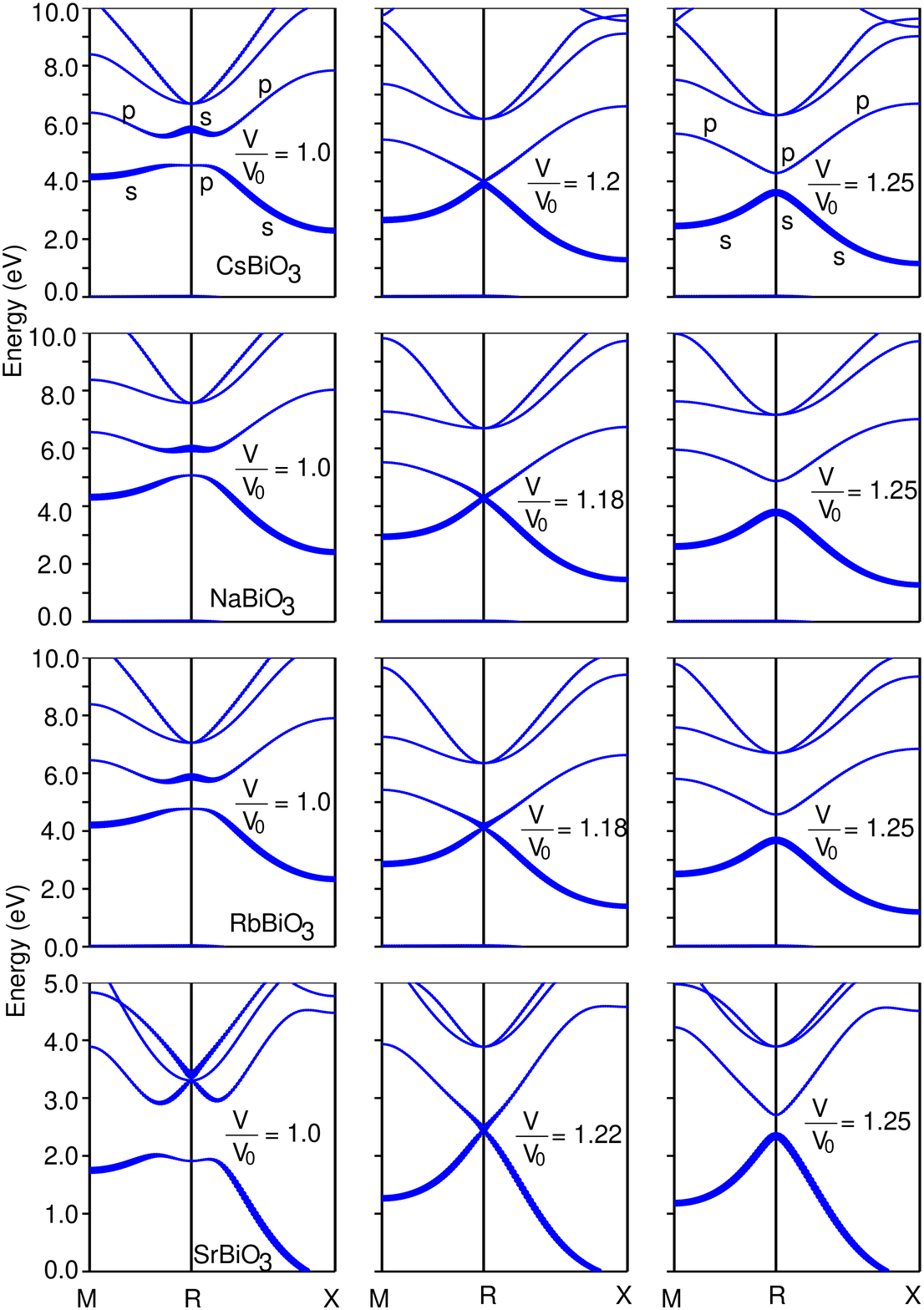}
\caption{The band structure of ABiO$_3$ as a function of volume expansion. The orbital contributions, around the high symmetry point $R$, to the lower two bands are indicated in the upper panel. With expansion, the s-p band inversion vanishes to create a normal band gap. }
\label{fig5s}
\end{figure}

\section{Effect of Exchange Correlation functionals on NI-TI Phase Transition}

The hydrostatic pressure required to drive the NI-TI phase transition solely depend on value of the band gap. However, the band gap values of the halide perovskites, as shown in  Table \ref{T1} are highly sensitive to the exchange correlation functionals used in the calculations. This makes it difficult to have a quantitative estimation of the critical pressure values at which the transition occurs.

To  find  a  plausible  solution  to  this  issue,  we  have  examined  the  band gap variation as a function of lattice parameter for three different exchange-correlation functionals, namely, GGA-PBE, GGA+mBJ and HSE06. For the case  of  CsSnCl$_3$ and  CsSnBr$_3$, the results are plotted in Fig. \ref{fig6s}. Interestingly, we find that, the slopes are nearly same for all the three exchange-correlation functionals.  This suggests that if the accurate band gap is known (by some means: experimental or theoretical) for a given lattice parameter, then the band gap for a given pressure, within an acceptable error limit, can be generated through extrapolation.  Hence, the critical pressures at which the transitions occur can be estimated with reasonably accuracy.

\begin{figure}[h]
\centering
\includegraphics[angle=0.0,origin=c,height=5cm,width=10cm]{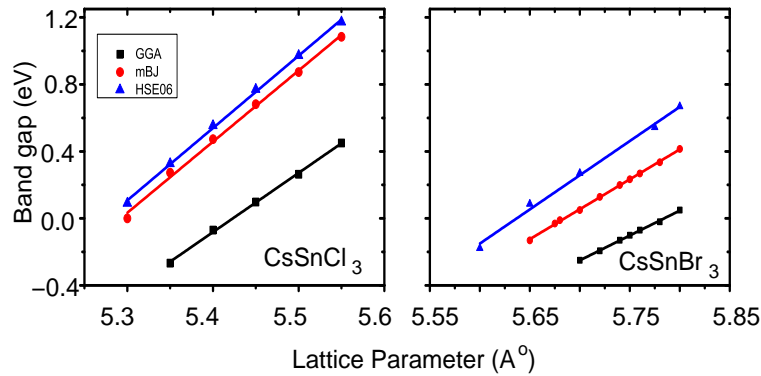}
\caption{Exchange-correlation functional dependent band gap of CsSnCl$_3$ and CsSnBr$_3$ as a function of  lattice parameter.  The slopes are found to be nearly same for each functionals. The HSE06 results are obtained using pseudpotential method through VASP simulation package\cite{Kresse,Joubert}.}
\label{fig6s}
\end{figure}

\end{widetext}
\bibliography{paper}

\begin{thebibliography}{43}%
\makeatletter
\providecommand \@ifxundefined [1]{%
 \@ifx{#1\undefined}
}%
\providecommand \@ifnum [1]{%
 \ifnum #1\expandafter \@firstoftwo
 \else \expandafter \@secondoftwo
 \fi
}%
\providecommand \@ifx [1]{%
 \ifx #1\expandafter \@firstoftwo
 \else \expandafter \@secondoftwo
 \fi
}%
\providecommand \natexlab [1]{#1}%
\providecommand \enquote  [1]{``#1''}%
\providecommand \bibnamefont  [1]{#1}%
\providecommand \bibfnamefont [1]{#1}%
\providecommand \citenamefont [1]{#1}%
\providecommand \href@noop [0]{\@secondoftwo}%
\providecommand \href [0]{\begingroup \@sanitize@url \@href}%
\providecommand \@href[1]{\@@startlink{#1}\@@href}%
\providecommand \@@href[1]{\endgroup#1\@@endlink}%
\providecommand \@sanitize@url [0]{\catcode `\\12\catcode `\$12\catcode
  `\&12\catcode `\#12\catcode `\^12\catcode `\_12\catcode `\%12\relax}%
\providecommand \@@startlink[1]{}%
\providecommand \@@endlink[0]{}%
\providecommand \url  [0]{\begingroup\@sanitize@url \@url }%
\providecommand \@url [1]{\endgroup\@href {#1}{\urlprefix }}%
\providecommand \urlprefix  [0]{URL }%
\providecommand \Eprint [0]{\href }%
\providecommand \doibase [0]{http://dx.doi.org/}%
\providecommand \selectlanguage [0]{\@gobble}%
\providecommand \bibinfo  [0]{\@secondoftwo}%
\providecommand \bibfield  [0]{\@secondoftwo}%
\providecommand \translation [1]{[#1]}%
\providecommand \BibitemOpen [0]{}%
\providecommand \bibitemStop [0]{}%
\providecommand \bibitemNoStop [0]{.\EOS\space}%
\providecommand \EOS [0]{\spacefactor3000\relax}%
\providecommand \BibitemShut  [1]{\csname bibitem#1\endcsname}%
\let\auto@bib@innerbib\@empty
\bibitem [{\citenamefont {Bernevig}\ \emph {et~al.}(2006)\citenamefont
  {Bernevig}, \citenamefont {Hughes},\ and\ \citenamefont {Zhang}}]{HgTe1}%
  \BibitemOpen
  \bibfield  {author} {\bibinfo {author} {\bibfnamefont {B.~A.}\ \bibnamefont
  {Bernevig}}, \bibinfo {author} {\bibfnamefont {T.~L.}\ \bibnamefont
  {Hughes}}, \ and\ \bibinfo {author} {\bibfnamefont {S.-C.}\ \bibnamefont
  {Zhang}},\ }\href {\doibase 10.1126/science.1133734} {\bibfield  {journal}
  {\bibinfo  {journal} {Science}\ }\textbf {\bibinfo {volume} {314}},\ \bibinfo
  {pages} {1757} (\bibinfo {year} {2006})}\BibitemShut {NoStop}%
\bibitem [{\citenamefont {K{\"o}nig}\ \emph {et~al.}(2007)\citenamefont
  {K{\"o}nig}, \citenamefont {Wiedmann}, \citenamefont {Br{\"u}ne},
  \citenamefont {Roth}, \citenamefont {Buhmann}, \citenamefont {Molenkamp},
  \citenamefont {Qi},\ and\ \citenamefont {Zhang}}]{HgTe2}%
  \BibitemOpen
  \bibfield  {author} {\bibinfo {author} {\bibfnamefont {M.}~\bibnamefont
  {K{\"o}nig}}, \bibinfo {author} {\bibfnamefont {S.}~\bibnamefont {Wiedmann}},
  \bibinfo {author} {\bibfnamefont {C.}~\bibnamefont {Br{\"u}ne}}, \bibinfo
  {author} {\bibfnamefont {A.}~\bibnamefont {Roth}}, \bibinfo {author}
  {\bibfnamefont {H.}~\bibnamefont {Buhmann}}, \bibinfo {author} {\bibfnamefont
  {L.~W.}\ \bibnamefont {Molenkamp}}, \bibinfo {author} {\bibfnamefont {X.-L.}\
  \bibnamefont {Qi}}, \ and\ \bibinfo {author} {\bibfnamefont {S.-C.}\
  \bibnamefont {Zhang}},\ }\href {\doibase 10.1126/science.1148047} {\bibfield
  {journal} {\bibinfo  {journal} {Science}\ }\textbf {\bibinfo {volume}
  {318}},\ \bibinfo {pages} {766} (\bibinfo {year} {2007})}\BibitemShut
  {NoStop}%
\bibitem [{\citenamefont {Hsieh}\ \emph {et~al.}(2008)\citenamefont {Hsieh},
  \citenamefont {Qian}, \citenamefont {Wray}, \citenamefont {Xia},
  \citenamefont {Hor}, \citenamefont {Cava},\ and\ \citenamefont
  {Hasan}}]{Bisb1}%
  \BibitemOpen
  \bibfield  {author} {\bibinfo {author} {\bibfnamefont {D.}~\bibnamefont
  {Hsieh}}, \bibinfo {author} {\bibfnamefont {D.}~\bibnamefont {Qian}},
  \bibinfo {author} {\bibfnamefont {L.}~\bibnamefont {Wray}}, \bibinfo {author}
  {\bibfnamefont {Y.}~\bibnamefont {Xia}}, \bibinfo {author} {\bibfnamefont
  {Y.~S.}\ \bibnamefont {Hor}}, \bibinfo {author} {\bibfnamefont {R.~J.}\
  \bibnamefont {Cava}}, \ and\ \bibinfo {author} {\bibfnamefont {M.~Z.}\
  \bibnamefont {Hasan}},\ }\href {\doibase 10.1038/nature06843} {\bibfield
  {journal} {\bibinfo  {journal} {Nature}\ }\textbf {\bibinfo {volume} {452}},\
  \bibinfo {pages} {970} (\bibinfo {year} {2008})},\ \Eprint
  {http://arxiv.org/abs/0910.2420} {0910.2420} \BibitemShut {NoStop}%
\bibitem [{\citenamefont {Hsieh}\ \emph
  {et~al.}(2009{\natexlab{a}})\citenamefont {Hsieh}, \citenamefont {Xia},
  \citenamefont {Wray}, \citenamefont {Qian}, \citenamefont {Pal},
  \citenamefont {Dil}, \citenamefont {Osterwalder}, \citenamefont {Meier},
  \citenamefont {Bihlmayer}, \citenamefont {Kane}, \citenamefont {Hor},
  \citenamefont {Cava},\ and\ \citenamefont {Hasan}}]{BiSb2}%
  \BibitemOpen
  \bibfield  {author} {\bibinfo {author} {\bibfnamefont {D.}~\bibnamefont
  {Hsieh}}, \bibinfo {author} {\bibfnamefont {Y.}~\bibnamefont {Xia}}, \bibinfo
  {author} {\bibfnamefont {L.}~\bibnamefont {Wray}}, \bibinfo {author}
  {\bibfnamefont {D.}~\bibnamefont {Qian}}, \bibinfo {author} {\bibfnamefont
  {A.}~\bibnamefont {Pal}}, \bibinfo {author} {\bibfnamefont {J.~H.}\
  \bibnamefont {Dil}}, \bibinfo {author} {\bibfnamefont {J.}~\bibnamefont
  {Osterwalder}}, \bibinfo {author} {\bibfnamefont {F.}~\bibnamefont {Meier}},
  \bibinfo {author} {\bibfnamefont {G.}~\bibnamefont {Bihlmayer}}, \bibinfo
  {author} {\bibfnamefont {C.~L.}\ \bibnamefont {Kane}}, \bibinfo {author}
  {\bibfnamefont {Y.~S.}\ \bibnamefont {Hor}}, \bibinfo {author} {\bibfnamefont
  {R.~J.}\ \bibnamefont {Cava}}, \ and\ \bibinfo {author} {\bibfnamefont
  {M.~Z.}\ \bibnamefont {Hasan}},\ }\href {\doibase 10.1126/science.1167733}
  {\bibfield  {journal} {\bibinfo  {journal} {Science}\ }\textbf {\bibinfo
  {volume} {323}},\ \bibinfo {pages} {919} (\bibinfo {year}
  {2009}{\natexlab{a}})}\BibitemShut {NoStop}%
\bibitem [{\citenamefont {Teo}\ \emph {et~al.}(2008)\citenamefont {Teo},
  \citenamefont {Fu},\ and\ \citenamefont {Kane}}]{BiSb3}%
  \BibitemOpen
  \bibfield  {author} {\bibinfo {author} {\bibfnamefont {J.~C.~Y.}\
  \bibnamefont {Teo}}, \bibinfo {author} {\bibfnamefont {L.}~\bibnamefont
  {Fu}}, \ and\ \bibinfo {author} {\bibfnamefont {C.~L.}\ \bibnamefont
  {Kane}},\ }\href {\doibase 10.1103/PhysRevB.78.045426} {\bibfield  {journal}
  {\bibinfo  {journal} {Phys. Rev. B}\ }\textbf {\bibinfo {volume} {78}},\
  \bibinfo {pages} {045426} (\bibinfo {year} {2008})}\BibitemShut {NoStop}%
\bibitem [{\citenamefont {Yang}\ \emph {et~al.}(2012)\citenamefont {Yang},
  \citenamefont {Setyawan}, \citenamefont {Wang}, \citenamefont {{Buongiorno
  Nardelli}},\ and\ \citenamefont {Curtarolo}}]{nature}%
  \BibitemOpen
  \bibfield  {author} {\bibinfo {author} {\bibfnamefont {K.}~\bibnamefont
  {Yang}}, \bibinfo {author} {\bibfnamefont {W.}~\bibnamefont {Setyawan}},
  \bibinfo {author} {\bibfnamefont {S.}~\bibnamefont {Wang}}, \bibinfo {author}
  {\bibfnamefont {M.}~\bibnamefont {{Buongiorno Nardelli}}}, \ and\ \bibinfo
  {author} {\bibfnamefont {S.}~\bibnamefont {Curtarolo}},\ }\href {\doibase
  10.1038/nmat3332} {\bibfield  {journal} {\bibinfo  {journal} {Nature
  Materials}\ }\textbf {\bibinfo {volume} {11}},\ \bibinfo {pages} {614}
  (\bibinfo {year} {2012})}\BibitemShut {NoStop}%
\bibitem [{\citenamefont {Zhang}\ \emph {et~al.}(2009)\citenamefont {Zhang},
  \citenamefont {Liu}, \citenamefont {Qi}, \citenamefont {Dai}, \citenamefont
  {Fang},\ and\ \citenamefont {Zhang}}]{bise1}%
  \BibitemOpen
  \bibfield  {author} {\bibinfo {author} {\bibfnamefont {H.}~\bibnamefont
  {Zhang}}, \bibinfo {author} {\bibfnamefont {C.-X.}\ \bibnamefont {Liu}},
  \bibinfo {author} {\bibfnamefont {X.-L.}\ \bibnamefont {Qi}}, \bibinfo
  {author} {\bibfnamefont {X.}~\bibnamefont {Dai}}, \bibinfo {author}
  {\bibfnamefont {Z.}~\bibnamefont {Fang}}, \ and\ \bibinfo {author}
  {\bibfnamefont {S.-C.}\ \bibnamefont {Zhang}},\ }\href {\doibase
  10.1038/nphys1270} {\bibfield  {journal} {\bibinfo  {journal} {Nature
  Physics}\ }\textbf {\bibinfo {volume} {5}},\ \bibinfo {pages} {438} (\bibinfo
  {year} {2009})}\BibitemShut {NoStop}%
\bibitem [{\citenamefont {Chen}\ \emph {et~al.}(2009)\citenamefont {Chen},
  \citenamefont {Analytis}, \citenamefont {Chu}, \citenamefont {Liu},
  \citenamefont {Mo}, \citenamefont {Qi}, \citenamefont {Zhang}, \citenamefont
  {Lu}, \citenamefont {Dai}, \citenamefont {Fang}, \citenamefont {Zhang},
  \citenamefont {Fisher}, \citenamefont {Hussain},\ and\ \citenamefont
  {Shen}}]{bise2}%
  \BibitemOpen
  \bibfield  {author} {\bibinfo {author} {\bibfnamefont {Y.~L.}\ \bibnamefont
  {Chen}}, \bibinfo {author} {\bibfnamefont {J.~G.}\ \bibnamefont {Analytis}},
  \bibinfo {author} {\bibfnamefont {J.-H.}\ \bibnamefont {Chu}}, \bibinfo
  {author} {\bibfnamefont {Z.~K.}\ \bibnamefont {Liu}}, \bibinfo {author}
  {\bibfnamefont {S.-K.}\ \bibnamefont {Mo}}, \bibinfo {author} {\bibfnamefont
  {X.~L.}\ \bibnamefont {Qi}}, \bibinfo {author} {\bibfnamefont {H.~J.}\
  \bibnamefont {Zhang}}, \bibinfo {author} {\bibfnamefont {D.~H.}\ \bibnamefont
  {Lu}}, \bibinfo {author} {\bibfnamefont {X.}~\bibnamefont {Dai}}, \bibinfo
  {author} {\bibfnamefont {Z.}~\bibnamefont {Fang}}, \bibinfo {author}
  {\bibfnamefont {S.~C.}\ \bibnamefont {Zhang}}, \bibinfo {author}
  {\bibfnamefont {I.~R.}\ \bibnamefont {Fisher}}, \bibinfo {author}
  {\bibfnamefont {Z.}~\bibnamefont {Hussain}}, \ and\ \bibinfo {author}
  {\bibfnamefont {Z.-X.}\ \bibnamefont {Shen}},\ }\href {\doibase
  10.1126/science.1173034} {\bibfield  {journal} {\bibinfo  {journal}
  {Science}\ }\textbf {\bibinfo {volume} {325}},\ \bibinfo {pages} {178}
  (\bibinfo {year} {2009})}\BibitemShut {NoStop}%
\bibitem [{\citenamefont {Xia}\ \emph {et~al.}(2009)\citenamefont {Xia},
  \citenamefont {Qian}, \citenamefont {Hsieh}, \citenamefont {Wray},
  \citenamefont {Pal}, \citenamefont {Lin}, \citenamefont {Bansil},
  \citenamefont {Grauer}, \citenamefont {Hor}, \citenamefont {Cava},\ and\
  \citenamefont {Hasan}}]{bise3}%
  \BibitemOpen
  \bibfield  {author} {\bibinfo {author} {\bibfnamefont {Y.}~\bibnamefont
  {Xia}}, \bibinfo {author} {\bibfnamefont {D.}~\bibnamefont {Qian}}, \bibinfo
  {author} {\bibfnamefont {D.}~\bibnamefont {Hsieh}}, \bibinfo {author}
  {\bibfnamefont {L.}~\bibnamefont {Wray}}, \bibinfo {author} {\bibfnamefont
  {A.}~\bibnamefont {Pal}}, \bibinfo {author} {\bibfnamefont {H.}~\bibnamefont
  {Lin}}, \bibinfo {author} {\bibfnamefont {A.}~\bibnamefont {Bansil}},
  \bibinfo {author} {\bibfnamefont {D.}~\bibnamefont {Grauer}}, \bibinfo
  {author} {\bibfnamefont {Y.~S.}\ \bibnamefont {Hor}}, \bibinfo {author}
  {\bibfnamefont {R.~J.}\ \bibnamefont {Cava}}, \ and\ \bibinfo {author}
  {\bibfnamefont {M.~Z.}\ \bibnamefont {Hasan}},\ }\href {\doibase
  10.1038/nphys1274} {\bibfield  {journal} {\bibinfo  {journal} {Nature
  Physics}\ }\textbf {\bibinfo {volume} {5}},\ \bibinfo {pages} {398} (\bibinfo
  {year} {2009})}\BibitemShut {NoStop}%
\bibitem [{\citenamefont {Hsieh}\ \emph
  {et~al.}(2009{\natexlab{b}})\citenamefont {Hsieh}, \citenamefont {Xia},
  \citenamefont {Qian}, \citenamefont {Wray}, \citenamefont {Meier},
  \citenamefont {Dil}, \citenamefont {Osterwalder}, \citenamefont {Patthey},
  \citenamefont {Fedorov}, \citenamefont {Lin}, \citenamefont {Bansil},
  \citenamefont {Grauer}, \citenamefont {Hor}, \citenamefont {Cava},\ and\
  \citenamefont {Hasan}}]{bise4}%
  \BibitemOpen
  \bibfield  {author} {\bibinfo {author} {\bibfnamefont {D.}~\bibnamefont
  {Hsieh}}, \bibinfo {author} {\bibfnamefont {Y.}~\bibnamefont {Xia}}, \bibinfo
  {author} {\bibfnamefont {D.}~\bibnamefont {Qian}}, \bibinfo {author}
  {\bibfnamefont {L.}~\bibnamefont {Wray}}, \bibinfo {author} {\bibfnamefont
  {F.}~\bibnamefont {Meier}}, \bibinfo {author} {\bibfnamefont {J.~H.}\
  \bibnamefont {Dil}}, \bibinfo {author} {\bibfnamefont {J.}~\bibnamefont
  {Osterwalder}}, \bibinfo {author} {\bibfnamefont {L.}~\bibnamefont
  {Patthey}}, \bibinfo {author} {\bibfnamefont {A.~V.}\ \bibnamefont
  {Fedorov}}, \bibinfo {author} {\bibfnamefont {H.}~\bibnamefont {Lin}},
  \bibinfo {author} {\bibfnamefont {A.}~\bibnamefont {Bansil}}, \bibinfo
  {author} {\bibfnamefont {D.}~\bibnamefont {Grauer}}, \bibinfo {author}
  {\bibfnamefont {Y.~S.}\ \bibnamefont {Hor}}, \bibinfo {author} {\bibfnamefont
  {R.~J.}\ \bibnamefont {Cava}}, \ and\ \bibinfo {author} {\bibfnamefont
  {M.~Z.}\ \bibnamefont {Hasan}},\ }\href {\doibase
  10.1103/PhysRevLett.103.146401} {\bibfield  {journal} {\bibinfo  {journal}
  {Phys. Rev. Lett.}\ }\textbf {\bibinfo {volume} {103}},\ \bibinfo {pages}
  {146401} (\bibinfo {year} {2009}{\natexlab{b}})}\BibitemShut {NoStop}%
\bibitem [{\citenamefont {Huang}\ and\ \citenamefont {Lambrecht}(2013)}]{ABX1}%
  \BibitemOpen
  \bibfield  {author} {\bibinfo {author} {\bibfnamefont {L.-y.}\ \bibnamefont
  {Huang}}\ and\ \bibinfo {author} {\bibfnamefont {W.~R.~L.}\ \bibnamefont
  {Lambrecht}},\ }\href {\doibase 10.1103/PhysRevB.88.165203} {\bibfield
  {journal} {\bibinfo  {journal} {Phys. Rev. B}\ }\textbf {\bibinfo {volume}
  {88}},\ \bibinfo {pages} {165203} (\bibinfo {year} {2013})}\BibitemShut
  {NoStop}%
\bibitem [{\citenamefont {Huang}\ and\ \citenamefont {Lambrecht}(2016)}]{ABX2}%
  \BibitemOpen
  \bibfield  {author} {\bibinfo {author} {\bibfnamefont {L.-y.}\ \bibnamefont
  {Huang}}\ and\ \bibinfo {author} {\bibfnamefont {W.~R.~L.}\ \bibnamefont
  {Lambrecht}},\ }\href {\doibase 10.1103/PhysRevB.93.195211} {\bibfield
  {journal} {\bibinfo  {journal} {Phys. Rev. B}\ }\textbf {\bibinfo {volume}
  {93}},\ \bibinfo {pages} {195211} (\bibinfo {year} {2016})}\BibitemShut
  {NoStop}%
\bibitem [{\citenamefont {Jin}\ \emph {et~al.}(2012)\citenamefont {Jin},
  \citenamefont {Im},\ and\ \citenamefont {Freeman}}]{ABX3}%
  \BibitemOpen
  \bibfield  {author} {\bibinfo {author} {\bibfnamefont {H.}~\bibnamefont
  {Jin}}, \bibinfo {author} {\bibfnamefont {J.}~\bibnamefont {Im}}, \ and\
  \bibinfo {author} {\bibfnamefont {A.~J.}\ \bibnamefont {Freeman}},\ }\href
  {\doibase 10.1103/PhysRevB.86.121102} {\bibfield  {journal} {\bibinfo
  {journal} {Phys. Rev. B}\ }\textbf {\bibinfo {volume} {86}},\ \bibinfo
  {pages} {121102} (\bibinfo {year} {2012})}\BibitemShut {NoStop}%
\bibitem [{\citenamefont {Afsari}\ \emph {et~al.}(2017)\citenamefont {Afsari},
  \citenamefont {Boochani}, \citenamefont {Hantezadeh},\ and\ \citenamefont
  {Elahi}}]{ABX4}%
  \BibitemOpen
  \bibfield  {author} {\bibinfo {author} {\bibfnamefont {M.}~\bibnamefont
  {Afsari}}, \bibinfo {author} {\bibfnamefont {A.}~\bibnamefont {Boochani}},
  \bibinfo {author} {\bibfnamefont {M.}~\bibnamefont {Hantezadeh}}, \ and\
  \bibinfo {author} {\bibfnamefont {S.~M.}\ \bibnamefont {Elahi}},\ }\href
  {\doibase 10.1016/j.ssc.2017.04.014} {\bibfield  {journal} {\bibinfo
  {journal} {Solid State Communications}\ }\textbf {\bibinfo {volume} {259}},\
  \bibinfo {pages} {10} (\bibinfo {year} {2017})}\BibitemShut {NoStop}%
\bibitem [{\citenamefont {Liu}\ \emph {et~al.}(2016)\citenamefont {Liu},
  \citenamefont {Kim}, \citenamefont {Tan},\ and\ \citenamefont
  {Rappe}}]{ABX5}%
  \BibitemOpen
  \bibfield  {author} {\bibinfo {author} {\bibfnamefont {S.}~\bibnamefont
  {Liu}}, \bibinfo {author} {\bibfnamefont {Y.}~\bibnamefont {Kim}}, \bibinfo
  {author} {\bibfnamefont {L.~Z.}\ \bibnamefont {Tan}}, \ and\ \bibinfo
  {author} {\bibfnamefont {A.~M.}\ \bibnamefont {Rappe}},\ }\href {\doibase
  10.1021/acs.nanolett.5b04545} {\bibfield  {journal} {\bibinfo  {journal}
  {Nano Letters}\ }\textbf {\bibinfo {volume} {16}},\ \bibinfo {pages} {1663}
  (\bibinfo {year} {2016})},\ \Eprint {http://arxiv.org/abs/1508.05625}
  {1508.05625} \BibitemShut {NoStop}%
\bibitem [{\citenamefont {Uchida}\ and\ \citenamefont
  {Kawasaki}(2018)}]{TI-Interface}%
  \BibitemOpen
  \bibfield  {author} {\bibinfo {author} {\bibfnamefont {M.}~\bibnamefont
  {Uchida}}\ and\ \bibinfo {author} {\bibfnamefont {M.}~\bibnamefont
  {Kawasaki}},\ }\href {http://stacks.iop.org/0022-3727/51/i=14/a=143001}
  {\bibfield  {journal} {\bibinfo  {journal} {Journal of Physics D: Applied
  Physics}\ }\textbf {\bibinfo {volume} {51}},\ \bibinfo {pages} {143001}
  (\bibinfo {year} {2018})}\BibitemShut {NoStop}%
\bibitem [{\citenamefont {Yang}\ and\ \citenamefont {Nagaosa}(2014)}]{DSM1}%
  \BibitemOpen
  \bibfield  {author} {\bibinfo {author} {\bibfnamefont {B.~J.}\ \bibnamefont
  {Yang}}\ and\ \bibinfo {author} {\bibfnamefont {N.}~\bibnamefont {Nagaosa}},\
  }\href {\doibase 10.1038/ncomms5898} {\bibfield  {journal} {\bibinfo
  {journal} {Nature Communications}\ }\textbf {\bibinfo {volume} {5}},\
  \bibinfo {pages} {1} (\bibinfo {year} {2014})},\ \Eprint
  {http://arxiv.org/abs/1404.0754} {1404.0754} \BibitemShut {NoStop}%
\bibitem [{\citenamefont {Sklyadneva}\ \emph {et~al.}(2016)\citenamefont
  {Sklyadneva}, \citenamefont {Rusinov}, \citenamefont {Heid}, \citenamefont
  {Bohnen}, \citenamefont {Echenique},\ and\ \citenamefont {Chulkov}}]{DSM2}%
  \BibitemOpen
  \bibfield  {author} {\bibinfo {author} {\bibfnamefont {I.~Y.}\ \bibnamefont
  {Sklyadneva}}, \bibinfo {author} {\bibfnamefont {I.~P.}\ \bibnamefont
  {Rusinov}}, \bibinfo {author} {\bibfnamefont {R.}~\bibnamefont {Heid}},
  \bibinfo {author} {\bibfnamefont {K.~P.}\ \bibnamefont {Bohnen}}, \bibinfo
  {author} {\bibfnamefont {P.~M.}\ \bibnamefont {Echenique}}, \ and\ \bibinfo
  {author} {\bibfnamefont {E.~V.}\ \bibnamefont {Chulkov}},\ }\href {\doibase
  10.1038/srep24137} {\bibfield  {journal} {\bibinfo  {journal} {Scientific
  Reports}\ }\textbf {\bibinfo {volume} {6}},\ \bibinfo {pages} {1} (\bibinfo
  {year} {2016})}\BibitemShut {NoStop}%
\bibitem [{\citenamefont {Yang}\ \emph {et~al.}(2017)\citenamefont {Yang},
  \citenamefont {Skelton}, \citenamefont {{Da Silva}}, \citenamefont {Frost},\
  and\ \citenamefont {Walsh}}]{crystal-cl}%
  \BibitemOpen
  \bibfield  {author} {\bibinfo {author} {\bibfnamefont {R.~X.}\ \bibnamefont
  {Yang}}, \bibinfo {author} {\bibfnamefont {J.~M.}\ \bibnamefont {Skelton}},
  \bibinfo {author} {\bibfnamefont {E.~L.}\ \bibnamefont {{Da Silva}}},
  \bibinfo {author} {\bibfnamefont {J.~M.}\ \bibnamefont {Frost}}, \ and\
  \bibinfo {author} {\bibfnamefont {A.}~\bibnamefont {Walsh}},\ }\href
  {\doibase 10.1021/acs.jpclett.7b02423} {\bibfield  {journal} {\bibinfo
  {journal} {Journal of Physical Chemistry Letters}\ }\textbf {\bibinfo
  {volume} {8}},\ \bibinfo {pages} {4720} (\bibinfo {year} {2017})},\ \Eprint
  {http://arxiv.org/abs/1708.00499} {1708.00499} \BibitemShut {NoStop}%
\bibitem [{\citenamefont {Scaife}\ \emph {et~al.}(1974)\citenamefont {Scaife},
  \citenamefont {Weller},\ and\ \citenamefont {Fisher}}]{crystal1}%
  \BibitemOpen
  \bibfield  {author} {\bibinfo {author} {\bibfnamefont {D.~E.}\ \bibnamefont
  {Scaife}}, \bibinfo {author} {\bibfnamefont {P.~F.}\ \bibnamefont {Weller}},
  \ and\ \bibinfo {author} {\bibfnamefont {W.~G.}\ \bibnamefont {Fisher}},\
  }\href {\doibase 10.1016/0022-4596(74)90088-7} {\bibfield  {journal}
  {\bibinfo  {journal} {Journal of Solid State Chemistry}\ }\textbf {\bibinfo
  {volume} {9}},\ \bibinfo {pages} {308} (\bibinfo {year} {1974})}\BibitemShut
  {NoStop}%
\bibitem [{\citenamefont {Peedikakkandy}\ and\ \citenamefont
  {Bhargava}(2016)}]{crystal2}%
  \BibitemOpen
  \bibfield  {author} {\bibinfo {author} {\bibfnamefont {L.}~\bibnamefont
  {Peedikakkandy}}\ and\ \bibinfo {author} {\bibfnamefont {P.}~\bibnamefont
  {Bhargava}},\ }\href {\doibase 10.1039/C5RA22317B} {\bibfield  {journal}
  {\bibinfo  {journal} {RSC Adv.}\ }\textbf {\bibinfo {volume} {6}},\ \bibinfo
  {pages} {19857} (\bibinfo {year} {2016})}\BibitemShut {NoStop}%
\bibitem [{\citenamefont {Hamann}(1979)}]{LAPW}%
  \BibitemOpen
  \bibfield  {author} {\bibinfo {author} {\bibfnamefont {D.~R.}\ \bibnamefont
  {Hamann}},\ }\href {\doibase 10.1103/PhysRevLett.42.662} {\bibfield
  {journal} {\bibinfo  {journal} {Phys. Rev. Lett.}\ }\textbf {\bibinfo
  {volume} {42}},\ \bibinfo {pages} {662} (\bibinfo {year} {1979})}\BibitemShut
  {NoStop}%
\bibitem [{\citenamefont {Blaha}\ \emph {et~al.}(2001)\citenamefont {Blaha},
  \citenamefont {Schwartz}, \citenamefont {Madsen}, \citenamefont {Kvasnicka},\
  and\ \citenamefont {Luitz}}]{Blaha}%
  \BibitemOpen
  \bibfield  {author} {\bibinfo {author} {\bibfnamefont {P.}~\bibnamefont
  {Blaha}}, \bibinfo {author} {\bibfnamefont {K.}~\bibnamefont {Schwartz}},
  \bibinfo {author} {\bibfnamefont {G.}~\bibnamefont {Madsen}}, \bibinfo
  {author} {\bibfnamefont {D.}~\bibnamefont {Kvasnicka}}, \ and\ \bibinfo
  {author} {\bibfnamefont {J.}~\bibnamefont {Luitz}},\ }\href@noop {} {\emph
  {\bibinfo {title} {WIEN2k An Augmanted Plane Wave+Local Orbitals Program for
  Calculating Crystal Properties}}}\ (\bibinfo  {publisher} {Karlheinz
  Schwartz, Tech. Universitt Wien, Austria},\ \bibinfo {year}
  {2001})\BibitemShut {NoStop}%
\bibitem [{\citenamefont {Perdew}\ \emph {et~al.}(1996)\citenamefont {Perdew},
  \citenamefont {Burke},\ and\ \citenamefont {Ernzerhof}}]{GGA}%
  \BibitemOpen
  \bibfield  {author} {\bibinfo {author} {\bibfnamefont {J.~P.}\ \bibnamefont
  {Perdew}}, \bibinfo {author} {\bibfnamefont {K.}~\bibnamefont {Burke}}, \
  and\ \bibinfo {author} {\bibfnamefont {M.}~\bibnamefont {Ernzerhof}},\ }\href
  {\doibase 10.1103/PhysRevLett.77.3865} {\bibfield  {journal} {\bibinfo
  {journal} {Phys. Rev. Lett.}\ }\textbf {\bibinfo {volume} {77}},\ \bibinfo
  {pages} {3865} (\bibinfo {year} {1996})}\BibitemShut {NoStop}%
\bibitem [{\citenamefont {Tran}\ and\ \citenamefont {Blaha}(2009)}]{mbj-1}%
  \BibitemOpen
  \bibfield  {author} {\bibinfo {author} {\bibfnamefont {F.}~\bibnamefont
  {Tran}}\ and\ \bibinfo {author} {\bibfnamefont {P.}~\bibnamefont {Blaha}},\
  }\href {\doibase 10.1103/PhysRevLett.102.226401} {\bibfield  {journal}
  {\bibinfo  {journal} {Phys. Rev. Lett.}\ }\textbf {\bibinfo {volume} {102}},\
  \bibinfo {pages} {226401} (\bibinfo {year} {2009})}\BibitemShut {NoStop}%
\bibitem [{\citenamefont {Tran}\ \emph {et~al.}(2007)\citenamefont {Tran},
  \citenamefont {Blaha},\ and\ \citenamefont {Schwarz}}]{mbj-2}%
  \BibitemOpen
  \bibfield  {author} {\bibinfo {author} {\bibfnamefont {F.}~\bibnamefont
  {Tran}}, \bibinfo {author} {\bibfnamefont {P.}~\bibnamefont {Blaha}}, \ and\
  \bibinfo {author} {\bibfnamefont {K.}~\bibnamefont {Schwarz}},\ }\href
  {http://stacks.iop.org/0953-8984/19/i=19/a=196208} {\bibfield  {journal}
  {\bibinfo  {journal} {Journal of Physics: Condensed Matter}\ }\textbf
  {\bibinfo {volume} {19}},\ \bibinfo {pages} {196208} (\bibinfo {year}
  {2007})}\BibitemShut {NoStop}%
\bibitem [{\citenamefont {Birch}(1947)}]{BM1}%
  \BibitemOpen
  \bibfield  {author} {\bibinfo {author} {\bibfnamefont {F.}~\bibnamefont
  {Birch}},\ }\href {\doibase 10.1103/PhysRev.71.809} {\bibfield  {journal}
  {\bibinfo  {journal} {Phys. Rev.}\ }\textbf {\bibinfo {volume} {71}},\
  \bibinfo {pages} {809} (\bibinfo {year} {1947})}\BibitemShut {NoStop}%
\bibitem [{\citenamefont {Murnaghan}(1944)}]{BM2}%
  \BibitemOpen
  \bibfield  {author} {\bibinfo {author} {\bibfnamefont {F.~D.}\ \bibnamefont
  {Murnaghan}},\ }\href@noop {} {\bibfield  {journal} {\bibinfo  {journal}
  {Proceedings of the National Academy of Sciences}\ }\textbf {\bibinfo
  {volume} {30}},\ \bibinfo {pages} {244} (\bibinfo {year} {1944})}\BibitemShut
  {NoStop}%
\bibitem [{\citenamefont {Kresse}\ and\ \citenamefont
  {Furthm\"uller}(1996)}]{Kresse}%
  \BibitemOpen
  \bibfield  {author} {\bibinfo {author} {\bibfnamefont {G.}~\bibnamefont
  {Kresse}}\ and\ \bibinfo {author} {\bibfnamefont {J.}~\bibnamefont
  {Furthm\"uller}},\ }\href {\doibase 10.1103/PhysRevB.54.11169} {\bibfield
  {journal} {\bibinfo  {journal} {Phys. Rev. B}\ }\textbf {\bibinfo {volume}
  {54}},\ \bibinfo {pages} {11169} (\bibinfo {year} {1996})}\BibitemShut
  {NoStop}%
\bibitem [{\citenamefont {Kresse}\ and\ \citenamefont
  {Joubert}(1999)}]{Joubert}%
  \BibitemOpen
  \bibfield  {author} {\bibinfo {author} {\bibfnamefont {G.}~\bibnamefont
  {Kresse}}\ and\ \bibinfo {author} {\bibfnamefont {D.}~\bibnamefont
  {Joubert}},\ }\href {\doibase 10.1103/PhysRevB.59.1758} {\bibfield  {journal}
  {\bibinfo  {journal} {Phys. Rev. B}\ }\textbf {\bibinfo {volume} {59}},\
  \bibinfo {pages} {1758} (\bibinfo {year} {1999})}\BibitemShut {NoStop}%
\bibitem [{\citenamefont {Tadano}\ \emph {et~al.}(2014)\citenamefont {Tadano},
  \citenamefont {Gohda},\ and\ \citenamefont {Tsuneyuki}}]{Tadano}%
  \BibitemOpen
  \bibfield  {author} {\bibinfo {author} {\bibfnamefont {T.}~\bibnamefont
  {Tadano}}, \bibinfo {author} {\bibfnamefont {Y.}~\bibnamefont {Gohda}}, \
  and\ \bibinfo {author} {\bibfnamefont {S.}~\bibnamefont {Tsuneyuki}},\
  }\href@noop {} {\bibfield  {journal} {\bibinfo  {journal} {Journal of
  Physics: Condensed Matter}\ }\textbf {\bibinfo {volume} {26}},\ \bibinfo
  {pages} {225402} (\bibinfo {year} {2014})}\BibitemShut {NoStop}%
\bibitem [{\citenamefont {Mostofi}\ \emph {et~al.}(2008)\citenamefont
  {Mostofi}, \citenamefont {Yates}, \citenamefont {Lee}, \citenamefont {Souza},
  \citenamefont {Vanderbilt},\ and\ \citenamefont {Marzari}}]{Mostofi}%
  \BibitemOpen
  \bibfield  {author} {\bibinfo {author} {\bibfnamefont {A.~A.}\ \bibnamefont
  {Mostofi}}, \bibinfo {author} {\bibfnamefont {J.~R.}\ \bibnamefont {Yates}},
  \bibinfo {author} {\bibfnamefont {Y.-S.}\ \bibnamefont {Lee}}, \bibinfo
  {author} {\bibfnamefont {I.}~\bibnamefont {Souza}}, \bibinfo {author}
  {\bibfnamefont {D.}~\bibnamefont {Vanderbilt}}, \ and\ \bibinfo {author}
  {\bibfnamefont {N.}~\bibnamefont {Marzari}},\ }\href@noop {} {\bibfield
  {journal} {\bibinfo  {journal} {Computer Physics Communications}\ }\textbf
  {\bibinfo {volume} {178}},\ \bibinfo {pages} {685} (\bibinfo {year}
  {2008})}\BibitemShut {NoStop}%
\bibitem [{\citenamefont {Wu}\ \emph {et~al.}(2018)\citenamefont {Wu},
  \citenamefont {Zhang}, \citenamefont {Song}, \citenamefont {Troyer},\ and\
  \citenamefont {Soluyanov}}]{arpes}%
  \BibitemOpen
  \bibfield  {author} {\bibinfo {author} {\bibfnamefont {Q.}~\bibnamefont
  {Wu}}, \bibinfo {author} {\bibfnamefont {S.}~\bibnamefont {Zhang}}, \bibinfo
  {author} {\bibfnamefont {H.-F.}\ \bibnamefont {Song}}, \bibinfo {author}
  {\bibfnamefont {M.}~\bibnamefont {Troyer}}, \ and\ \bibinfo {author}
  {\bibfnamefont {A.~A.}\ \bibnamefont {Soluyanov}},\ }\href {\doibase
  https://doi.org/10.1016/j.cpc.2017.09.033} {\bibfield  {journal} {\bibinfo
  {journal} {Computer Physics Communications}\ }\textbf {\bibinfo {volume}
  {224}},\ \bibinfo {pages} {405 } (\bibinfo {year} {2018})}\BibitemShut
  {NoStop}%
\bibitem [{\citenamefont {Guinea}\ \emph {et~al.}(1983)\citenamefont {Guinea},
  \citenamefont {Tejedor}, \citenamefont {Flores},\ and\ \citenamefont
  {Louis}}]{LDOS1}%
  \BibitemOpen
  \bibfield  {author} {\bibinfo {author} {\bibfnamefont {F.}~\bibnamefont
  {Guinea}}, \bibinfo {author} {\bibfnamefont {C.}~\bibnamefont {Tejedor}},
  \bibinfo {author} {\bibfnamefont {F.}~\bibnamefont {Flores}}, \ and\ \bibinfo
  {author} {\bibfnamefont {E.}~\bibnamefont {Louis}},\ }\href {\doibase
  10.1103/PhysRevB.28.4397} {\bibfield  {journal} {\bibinfo  {journal} {Phys.
  Rev. B}\ }\textbf {\bibinfo {volume} {28}},\ \bibinfo {pages} {4397}
  (\bibinfo {year} {1983})}\BibitemShut {NoStop}%
\bibitem [{\citenamefont {Sancho}\ \emph {et~al.}(1984)\citenamefont {Sancho},
  \citenamefont {Sancho},\ and\ \citenamefont {Rubio}}]{Sancho}%
  \BibitemOpen
  \bibfield  {author} {\bibinfo {author} {\bibfnamefont {M.~P.~L.}\
  \bibnamefont {Sancho}}, \bibinfo {author} {\bibfnamefont {J.~M.~L.}\
  \bibnamefont {Sancho}}, \ and\ \bibinfo {author} {\bibfnamefont
  {J.}~\bibnamefont {Rubio}},\ }\href@noop {} {\bibfield  {journal} {\bibinfo
  {journal} {Journal of Physics F: Metal Physics}\ }\textbf {\bibinfo {volume}
  {14}},\ \bibinfo {pages} {1205} (\bibinfo {year} {1984})}\BibitemShut
  {NoStop}%
\bibitem [{\citenamefont {Sancho}\ \emph {et~al.}(1985)\citenamefont {Sancho},
  \citenamefont {Sancho}, \citenamefont {Sancho},\ and\ \citenamefont
  {Rubio}}]{Lopez}%
  \BibitemOpen
  \bibfield  {author} {\bibinfo {author} {\bibfnamefont {M.~P.~L.}\
  \bibnamefont {Sancho}}, \bibinfo {author} {\bibfnamefont {J.~M.~L.}\
  \bibnamefont {Sancho}}, \bibinfo {author} {\bibfnamefont {J.~M.~L.}\
  \bibnamefont {Sancho}}, \ and\ \bibinfo {author} {\bibfnamefont
  {J.}~\bibnamefont {Rubio}},\ }\href@noop {} {\bibfield  {journal} {\bibinfo
  {journal} {Journal of Physics F: Metal Physics}\ }\textbf {\bibinfo {volume}
  {15}},\ \bibinfo {pages} {851} (\bibinfo {year} {1985})}\BibitemShut
  {NoStop}%
\bibitem [{\citenamefont {Khamari}\ \emph {et~al.}(2018)\citenamefont
  {Khamari}, \citenamefont {Kashikar},\ and\ \citenamefont {Nanda}}]{ABO-1}%
  \BibitemOpen
  \bibfield  {author} {\bibinfo {author} {\bibfnamefont {B.}~\bibnamefont
  {Khamari}}, \bibinfo {author} {\bibfnamefont {R.}~\bibnamefont {Kashikar}}, \
  and\ \bibinfo {author} {\bibfnamefont {B.~R.~K.}\ \bibnamefont {Nanda}},\
  }\href {\doibase 10.1103/PhysRevB.97.045149} {\bibfield  {journal} {\bibinfo
  {journal} {Phys. Rev. B}\ }\textbf {\bibinfo {volume} {97}},\ \bibinfo
  {pages} {045149} (\bibinfo {year} {2018})}\BibitemShut {NoStop}%
\bibitem [{\citenamefont {Yan}\ \emph {et~al.}(2013)\citenamefont {Yan},
  \citenamefont {Jansen},\ and\ \citenamefont {Felser}}]{ABO-2}%
  \BibitemOpen
  \bibfield  {author} {\bibinfo {author} {\bibfnamefont {B.}~\bibnamefont
  {Yan}}, \bibinfo {author} {\bibfnamefont {M.}~\bibnamefont {Jansen}}, \ and\
  \bibinfo {author} {\bibfnamefont {C.}~\bibnamefont {Felser}},\ }\href
  {http://dx.doi.org/10.1038/nphys2762 http://10.0.4.14/nphys2762
  https://www.nature.com/articles/nphys2762{\#}supplementary-information}
  {\bibfield  {journal} {\bibinfo  {journal} {Nature Physics}\ }\textbf
  {\bibinfo {volume} {9}},\ \bibinfo {pages} {709} (\bibinfo {year}
  {2013})}\BibitemShut {NoStop}%
\bibitem [{\citenamefont {Li}\ \emph {et~al.}(2015)\citenamefont {Li},
  \citenamefont {Yan}, \citenamefont {Thomale},\ and\ \citenamefont
  {Hanke}}]{ABO-3}%
  \BibitemOpen
  \bibfield  {author} {\bibinfo {author} {\bibfnamefont {G.}~\bibnamefont
  {Li}}, \bibinfo {author} {\bibfnamefont {B.}~\bibnamefont {Yan}}, \bibinfo
  {author} {\bibfnamefont {R.}~\bibnamefont {Thomale}}, \ and\ \bibinfo
  {author} {\bibfnamefont {W.}~\bibnamefont {Hanke}},\ }\href
  {http://dx.doi.org/10.1038/srep10435} {\bibfield  {journal} {\bibinfo
  {journal} {Scientific Reports}\ }\textbf {\bibinfo {volume} {5}},\ \bibinfo
  {pages} {10435} (\bibinfo {year} {2015})}\BibitemShut {NoStop}%
\bibitem [{\citenamefont {Zhang}\ \emph {et~al.}(2017)\citenamefont {Zhang},
  \citenamefont {Abdalla}, \citenamefont {Liu},\ and\ \citenamefont
  {Zunger}}]{ABX6}%
  \BibitemOpen
  \bibfield  {author} {\bibinfo {author} {\bibfnamefont {X.}~\bibnamefont
  {Zhang}}, \bibinfo {author} {\bibfnamefont {L.~B.}\ \bibnamefont {Abdalla}},
  \bibinfo {author} {\bibfnamefont {Q.}~\bibnamefont {Liu}}, \ and\ \bibinfo
  {author} {\bibfnamefont {A.}~\bibnamefont {Zunger}},\ }\href {\doibase
  10.1002/adfm.201701266} {\bibfield  {journal} {\bibinfo  {journal} {Advanced
  Functional Materials}\ }\textbf {\bibinfo {volume} {27}},\ \bibinfo {pages}
  {1} (\bibinfo {year} {2017})},\ \Eprint {http://arxiv.org/abs/1611.06839}
  {1611.06839} \BibitemShut {NoStop}%
\bibitem [{\citenamefont {Wojek}\ \emph {et~al.}(2014)\citenamefont {Wojek},
  \citenamefont {Dziawa}, \citenamefont {Kowalski}, \citenamefont
  {Szczerbakow}, \citenamefont {Black-Schaffer}, \citenamefont {Berntsen},
  \citenamefont {Balasubramanian}, \citenamefont {Story},\ and\ \citenamefont
  {Tjernberg}}]{negative1}%
  \BibitemOpen
  \bibfield  {author} {\bibinfo {author} {\bibfnamefont {B.~M.}\ \bibnamefont
  {Wojek}}, \bibinfo {author} {\bibfnamefont {P.}~\bibnamefont {Dziawa}},
  \bibinfo {author} {\bibfnamefont {B.~J.}\ \bibnamefont {Kowalski}}, \bibinfo
  {author} {\bibfnamefont {A.}~\bibnamefont {Szczerbakow}}, \bibinfo {author}
  {\bibfnamefont {A.~M.}\ \bibnamefont {Black-Schaffer}}, \bibinfo {author}
  {\bibfnamefont {M.~H.}\ \bibnamefont {Berntsen}}, \bibinfo {author}
  {\bibfnamefont {T.}~\bibnamefont {Balasubramanian}}, \bibinfo {author}
  {\bibfnamefont {T.}~\bibnamefont {Story}}, \ and\ \bibinfo {author}
  {\bibfnamefont {O.}~\bibnamefont {Tjernberg}},\ }\href {\doibase
  10.1103/PhysRevB.90.161202} {\bibfield  {journal} {\bibinfo  {journal} {Phys.
  Rev. B}\ }\textbf {\bibinfo {volume} {90}},\ \bibinfo {pages} {161202}
  (\bibinfo {year} {2014})}\BibitemShut {NoStop}%
\bibitem [{\citenamefont {Qi}\ and\ \citenamefont {Zhang}(2011)}]{negative2}%
  \BibitemOpen
  \bibfield  {author} {\bibinfo {author} {\bibfnamefont {X.-L.}\ \bibnamefont
  {Qi}}\ and\ \bibinfo {author} {\bibfnamefont {S.-C.}\ \bibnamefont {Zhang}},\
  }\href {\doibase 10.1103/RevModPhys.83.1057} {\bibfield  {journal} {\bibinfo
  {journal} {Rev. Mod. Phys.}\ }\textbf {\bibinfo {volume} {83}},\ \bibinfo
  {pages} {1057} (\bibinfo {year} {2011})}\BibitemShut {NoStop}%
\bibitem [{\citenamefont {Slater}\ and\ \citenamefont {Koster}(1954)}]{slater}%
  \BibitemOpen
  \bibfield  {author} {\bibinfo {author} {\bibfnamefont {J.~C.}\ \bibnamefont
  {Slater}}\ and\ \bibinfo {author} {\bibfnamefont {G.~F.}\ \bibnamefont
  {Koster}},\ }\href {\doibase 10.1103/PhysRev.94.1498} {\bibfield  {journal}
  {\bibinfo  {journal} {Phys. Rev.}\ }\textbf {\bibinfo {volume} {94}},\
  \bibinfo {pages} {1498} (\bibinfo {year} {1954})}\BibitemShut {NoStop}%
\end{thebibliography}

\end{document}